\documentclass{article}

\usepackage{arxiv}

\usepackage[utf8]{inputenc} 
\usepackage[T1]{fontenc}    
\usepackage{hyperref}       
\usepackage{url}            
\usepackage{booktabs}       
\usepackage{amsfonts}       
\usepackage{nicefrac}       
\usepackage{microtype}      

\usepackage{epstopdf}
\usepackage{physics}
\usepackage[font=small]{caption} 
\usepackage[backend=biber,style=ieee,citestyle=numeric-comp,natbib=true,sorting=none,sortcites=true,firstinits=true,url=false,eprint=false,labeldate=year]{biblatex}
    \newcommand\apptitle{\huge{Appendix}}
\usepackage{xifthen}
\usepackage{xparse}

\def\sp{\;\;\;\;}

\def\vec#1{\mbox{\boldmath $#1$} }

\newcommand{\scnr}[1]{(\textit{#1})}

\def\eqref#1{(\ref{#1})}
\def\eq#1{equation \eqref{#1}}
\def\eqs#1{equations \eqref{#1}}
\def\Eq#1{Equation \eqref{#1}}

\def\eqss#1#2{\eqs{#1} and \eqref{#2}}
\def\eqsss#1#2#3{\eqs{#1}, \eqref{#2} and \eqref{#3}}

\newcommand{\figsubref}[1]{\ifthenelse{\isempty{#1}}{}{(\textit{#1})}}
\newcommand{\figref}[2][]{\ref{#2}\figsubref{#1}}

\newcommand{\fig}[2][]{figure \ref{#2}\figsubref{#1}}
\newcommand{\figs}[2][]{figures \ref{#2}\figsubref{#1}}
\newcommand{\Fig}[2][]{Figure \ref{#2}\figsubref{#1}}
\newcommand{\Figs}[2][]{Figures \ref{#2}\figsubref{#1}}

\NewDocumentCommand{\figss}{O{} O{} m m}{\figs[#1]{#3} and \figref[#2]{#4}}
\NewDocumentCommand{\figsss}{O{} O{} O{} m m m}{\figs[#1]{#4}, \figref[#2]{#5} and \figref[#3]{#6}}
\NewDocumentCommand{\Figss}{O{} O{} m m}{\Figs[#1]{#3} and \figref[#2]{#4}}
\NewDocumentCommand{\Figsss}{O{} O{} O{} m m m}{\Figs[#1]{#4}, \figref[#2]{#5} and \figref[#3]{#6}}

\def\sec#1{\S \ref{#1}}
\def\secs#1{\S\S \ref{#1}}

\def\secss#1#2{\secs{#1} and \ref{#2}}
\def\secsss#1#2#3{\secs{#1}, \ref{#2} and \ref{#3}}

\def\app#1{appendix \ref{#1}}

\def\be{\begin{equation}}
\def\ee{\end{equation}}

\def\bea{\begin{eqnarray}}
\def\eea{\end{eqnarray}}

\def\bml{\begin{multline}}
\def\eml{\end{multline}}

\def\ba{\begin{align}}
\def\ea{\end{align}}

\newenvironment{packed_enumerate}{
\begin{enumerate}[(a)]
  \setlength{\itemsep}{1pt}
  \setlength{\parskip}{0pt}
  \setlength{\parsep}{0pt}
}{\end{enumerate}}

\def\ve{\vec{e}}

\def\vR{\vec{R}}

\def\vx{\vec{x}}

\def\vnabla{\vec{\nabla}}

\def\vnabla{\vec{\nabla}}

\def\ba{\overline{a}}

\usepackage{graphicx}
\title{Optimising branched fluidic networks: A unifying approach including laminar and turbulent flows, rough walls, and non-Newtonian fluids}

\author{
 J.S. Smink, R. Hagmeijer, C.H. Venner, and C.W. Visser \\
  Engineering Fluid Dynamics\\
  University of Twente\\
  7522 NB Enschede\\
  The Netherlands\\
  \texttt{j.s.smink@utwente.nl} }

\begin{document}
\maketitle
\begin{center}
\textsc{A preprint}\\
15 January 2025\\
\vspace{1cm}
\end{center}
\begin{abstract}
Power minimisation in branched fluidic networks has gained significant attention in biology and engineering. The optimal network is defined by channel radii that minimise the sum of viscous dissipation and the volumetric energetic cost of the fluid. For limit cases including laminar flows, high Reynolds number turbulence, or smooth channel approximations, optimal solutions are known. However, no single optimisation approach captures these limit cases. Furthermore, realistic fluidic networks exhibit intermediate points in the parameter space that can hardly be optimised. Here, we present a unifying optimisation approach based on the Darcy friction factor, which has been determined for a wide range of flow regimes and fluid models. We optimise fluidic networks for the entire parameter space:
\begin{itemize}
\item[-] Laminar and turbulent flows, including networks that exhibit both flow types,
\item[-] Non-Newtonian fluid models (power-law, Bingham, and Herschel-Bulkley), and
\item[-] Networks with arbitrary wall roughness, including non-uniform relative roughness.
\end{itemize}
The optimal channel radii are presented analytically and graphically. Finally, the parameter $x$ in the optimisation relationship $Q\propto R^x$ was approximated as a function of the Reynolds number, revealing that previously-determined values of $x$ hardly apply to realistic turbulent flows. Our approach can be extended to other configurations for which the friction factor is known, such as different channel curvatures or wall slip conditions, enabling optimisation of a wide range of fluidic networks.
\end{abstract}


\section{Introduction}

Branched networks for fluid transport are omnipresent in biology and engineering. Biological systems, such as vascular networks \citep{Hutchins1976,Kelch2015,Reichold2009} and the bronchial trees of the lungs \citep{Hooper1977,Xu2016,Sznitman2022} are capable of efficiently transporting heat or mass with low dissipation within limited volumes \citep{Bejan2013}. Similarly, fluidic networks provide efficient transport in emerging engineering applications, such as microfluidics \citep{Whitesides2006}, additive manufacturing \citep{Visser_foam,Shaqfeh2023}, creation of synthetic vasculatures \citep{Sexton2023}, microreactors \citep{Dong2021} and multifunctional materials \citep{Zheng2017}. These networks were comprehensively analysed and optimised for Newtonian \citep{Murray1926a,Murray1926c,Zamir1977, Kamiya_etal1974,Oka1987}, power-law \citep{Mayrovitz1987,Revellin_etal2009,Miguel2018, Stephenson_and_Lockerby2016}, and yield-stress fluids \citep{Ponalagusamy2012,Smink2023} in the laminar flow regime. However, a transition to turbulence commonly occurs in natural flows \citep{Olson1970,Calmet2016,Stein1976,Ku1997,Ha2018,Ha2019,Berger2000} and industrial flows such as district heating \citep{Gumpert2019,Steinegger2023}, water distribution networks, heat exchangers \citep{Siddiqui2017}, the paper making process \citep{Lundell2011}, and inertial microfluidics \citep{Wang2014}. Network optimisation is hardly possible for these systems, especially if rough channel walls or non-Newtonian fluid properties must be incorporated.

The parameter space spun by the Reynolds number, the roughness, and different fluid models is shown in \fig[a,b]{fig:overview}. Applications are plotted in \fig[a]{fig:overview}, showing the broad range of relevant Reynolds numbers $Re$ and relative channel wall roughness $\delta$, especially in the intermediate-$Re$ turbulent regime (the Reynolds number is defined in \eq{eq:main_def_Re}). \Fig[b]{fig:overview} shows that the energy consumption by the network has been minimised \citep{Murray1926a,Smink2023} only for limited parts of this parameter space. This total energy consists of the sum of the power needed to maintain the flow (product of the pressure drop $\Delta p$ and the volume flow rate $Q$) and a volumetric cost function $\alpha V$, with $\alpha$ a cost factor and $V$ the volume of the channel, for networks as schematically shown in \fig[c,d]{fig:overview}. For example, in the case of a vascular network, the cost factor expresses the metabolic energy needed to maintain blood ($\alpha\approx $ 1 kW m$^{-3}$) \citep{Murray1926a}; for other situations, see \citet{Smink2023}. Note that optimizing the relative channel radii for a given volume is a related problem, which has been pursued for many cases with e.g. constructional law \citep{Bejan2000,Miguel2018book,Miguel2018}. We focus on optimisation against a cost factor, which is generalisable for different boundary conditions such as a cost function of the wall surface area \citep{Woldenberg1986} and readily provides access to the optimal channel radii.

For a circular pipe with radius $R$, the power is generally minimised if: 
\begin{equation}
\frac{R^x}{Q} = \textnormal{const.}
\label{eq:def_x}
\end{equation}
with $x$ a flow-regime and fluid-model dependent power. $x = 3$ for laminar flow of an incompressible Newtonian fluid, as further analysed theoretically (e.g. revealing the shear forces, the optimal angles between channels, or special cases such as curved pipes or porous channels) \citep{Zamir1977, Sherman1981,Kamiya_etal1974,Miguel2018} and verified experimentally for e.g. human coronary and cerebral arteries \citep{Rossitti1993,Hutchins1976}. $x = 3$ also holds for laminar flows of different fluid models, including power-law  \citep{Mayrovitz1987,Revellin_etal2009,Miguel2018, Stephenson_and_Lockerby2016} and yield-stress fluids, such as the Bingham, Herschel Bulkley and Casson models \citep{Smink2023}, as well as for channels with elliptical \citep{Tesch2010} and rectangular \citep{Emerson2012} cross-sections.

For turbulent flows, fluidic networks were optimised only for limit cases and approximations \citep{Uylings1977,Bejan2000,Woldenberg1986, Stephenson_and_Lockerby2016}. For turbulent flow of Newtonian fluids at intermediate Reynolds numbers $2\times 10^4 <Re<10^6$ in a hydraulically smooth channel (shaded for relative wall roughness $\delta \rightarrow 0$ in \fig[b]{fig:overview}), $x = 17 / 7$ was derived from an empirical relation for the friction factor \citep{Bejan2013_convectionBook,Kou_etal2014}. For complete turbulence (the light-shaded area in \fig[b]{fig:overview}, which is defined as $Re>3500/\delta$ \citep{Moody1944}), it was derived that $x = 7 / 3$ \citep{Uylings1977,Stephenson_and_Lockerby2016, Bejan2000,Williams2007}. However, as we will show, this limit case is hardly reached in practice. Therefore, optimisation of fluidic networks that exhibit turbulent flows at intermediate Reynolds numbers is currently out of reach. Furthermore, existing approaches are so different that they are hardly comparable. Finally, a lack of model results for fluidic networks of intermediate complexity hampers validation of computational fluid dynamics (CFD), which is increasingly used to describe \citep{Morris2016} or design \citep{Sexton2023} networks with e.g. complex geometries.

\begin{figure}[h!]
\centering
   \includegraphics[width=\textwidth]{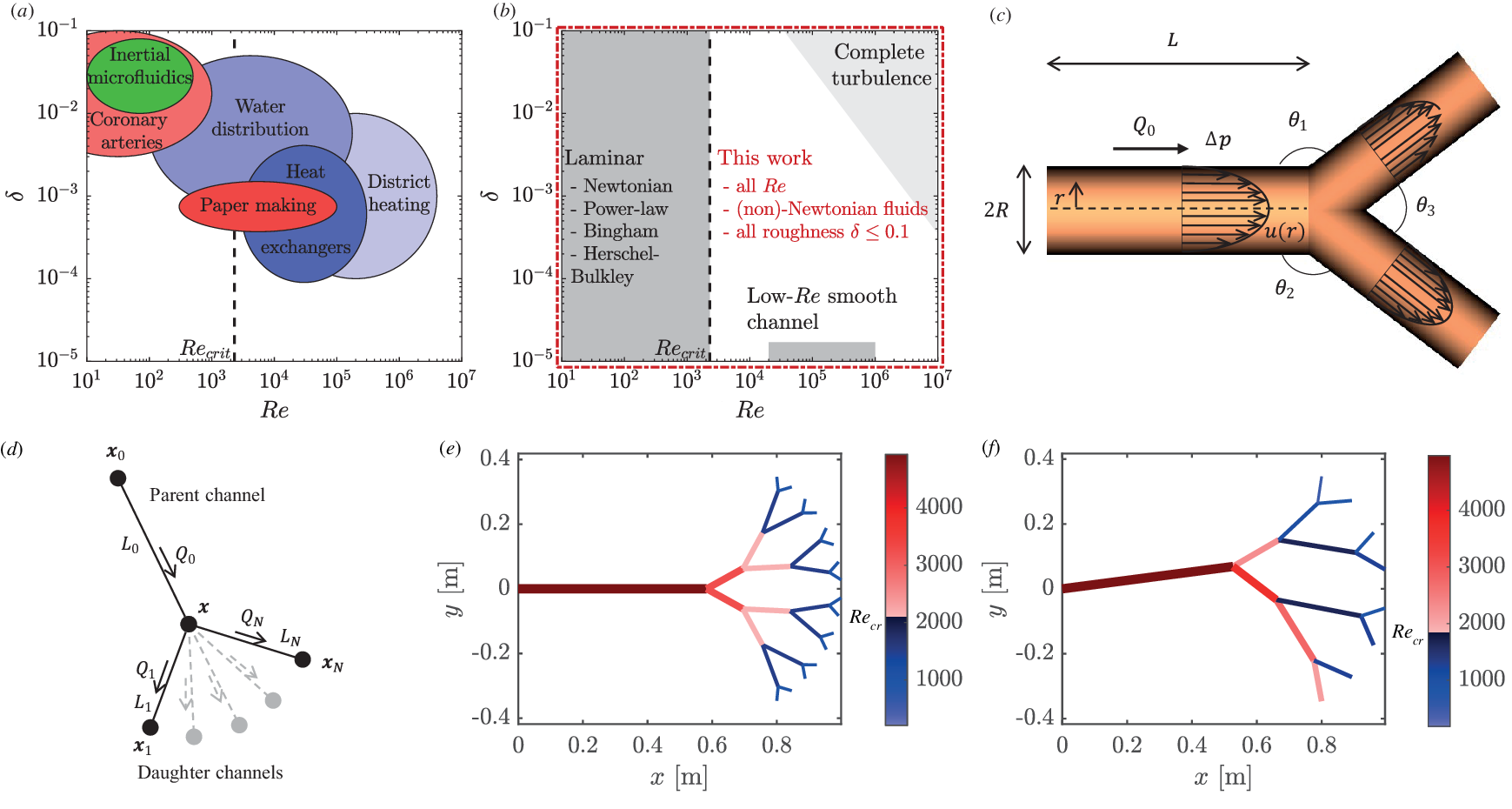} 
\caption{\scnr{a} Parameter space of the Reynolds number $Re$ and the relative channel wall roughness $\delta$ for fluidic networks in coronary arteries \citep{Burton2019,Kassab1993,Singhal1973}, paper making \citep{Grossman1968,Moller1976,Forgacs1957}, water distribution \citep{Energieverbruik-drinkwaterwinning2015,NEN1006+A1_2018}, inertial microfluidics \citep{Zhang2016,DiCarlo2009,Lu2017}, heat exchangers \citep{Towler2013book}, and district heating \citep{Gumpert2019,Steinegger2023}. The colour indicates a Newtonian (blue), power-law (green) or yield-stress (red) fluid. \scnr{b} Previously optimised parts of the parameter space include laminar flow and low-$Re$ flow in a smooth channel and a specific condition for complete rough-channel turbulence. \scnr{c} Schematic of a single branch with fully developed laminar flow profiles within the channels. \scnr{d} Schematic of a branched fluidic network, where a parent channel splits up into $N$ daughter channels. The location of the branching point $\vx$ follows from the analysis and determines the lengths $L_i$ of the channels. The grey channels indicate that it is possible to have many channels that originate from the branching point. \scnr{e,f} Examples of optimised branched fluidic networks. The colour indicates the Reynolds numbers in the channel. The begin and end-nodes are given as an input; the location of intermediate nodes follows from the optimisation. The following quantities are kept constant: $Q_0 = 20$ L min$^{-1}$, $\rho=1000$ kg m$^{-3}$, $\alpha=10^3$ W m$^{-3}$ and $\tau_0=0$ Pa. Fluid and system parameters are defined in \sec{sec:main_grote_sectie_optimalisatie}. \scnr{e} Newtonian fluid in rough channel ($\mu'=10^{-3}$ Pa s, $n=1.0$, $\varepsilon = 10^{-5}$ m, 5 levels, symmetric branching). \scnr{f} Power-law fluid in smooth channel ($\mu'=5\times 10^{-5}$ Pa s$^{1.5}$, $n=1.5$, 4 levels, asymmetric branching 1:2).}
\label{fig:overview}
\end{figure}

Therefore, in this study, we will introduce a universal approach for optimizing fluidic networks in the entire parameter space. The optimisation procedure is applicable to any flow regime, wall roughness, or fluid model for which the Darcy friction factor is known. First, \sec{sec:main_grote_sectie_optimalisatie} describes non-dimensionalisation of pipe flows, enabling calculation of the optimal Reynolds number within a channel and the corresponding channel radius. Subsequently, the optimal channel radius is derived for laminar and turbulent flow at arbitrary $Re$, arbitrary channel roughness $\delta$ and for both Newtonian and non-Newtonian fluids in \sec{sec:Flow_models}. \sec{sec:Synthesis} synthesises the results from \secss{sec:main_grote_sectie_optimalisatie}{sec:Flow_models}, resulting in approximations of $x$ as a function of $Re$ for all flow regimes and fluid models. The design procedure is presented in \sec{sec:designprocedure}. The conclusions are presented in \sec{sec:Turb_conclusion}. The design of the optimal location of branching points within the network is described in \app{sec:branching_point}, resulting in fully optimised networks for e.g. rough-wall channels (\fig[e]{fig:overview}) or for non-Newtonian fluids (\fig[f]{fig:overview}).


\section{Optimisation of fluidic branching}
\label{sec:main_grote_sectie_optimalisatie}

Consider a branching configuration comprising a parent channel connected to $N$ daughter channels at a branching point denoted as $\vx$, as depicted in \fig[d]{fig:overview}. The channels are labeled $0$ to $N$, with index $0$ indicating the parent channel. The effective radii of the channels are represented as $\vR\equiv(R_0, R_1,...,R_N)$, while the termination points of the channels are fixed and denoted as $\vx_i$ for $i=0,1,...,N$. The flow rates in the daughter channels are specified as $Q_i$ for $i=1,...,N$, with $Q_0$ corresponding to the flow rate in the parent channel. $Q_0$ is considered positive towards the branching point, whereas the flow rates in the daughter channels ($Q_i$, $i=1,...,N$) are taken positive away from the branching point. To ensure mass conservation, assuming incompressibility, the flow rates must satisfy $Q_0=\sum_{i=1}^N Q_i.$ The lengths of the channels, $L_i$, are functions of the branching location, given by 
$L_i\equiv|\vx_i-\vx|$ for  $i=0,1,...,N$. In this work, it is assumed that there is only an axial velocity $u$. The channels are assumed to be cylindrical with $R<<L$. The fluid properties are taken constant and the flow is fully developed.


For optimisation of a fluidic network, the power to maintain the flow and to maintain a fluid is minimised. The power is represented by a cost function depending on the radii and lengths of the channels, and is the sum of the individual channel contributions:
\be
P(\vR,\vx) \equiv \sum_{i=0}^N \left(\Delta p_i\, Q_i + \alpha V_i\right).
\label{eq:main branching power}
\ee
Here, $V_i \equiv \pi R_i^2 L_i$ is the channel volume and the pressure gradient $\frac{\Delta p}{L}$ is a fluid-type and flow regime dependent function. The pressure at the nodes is not defined or constrained. Differentiation of \eq{eq:main branching power} to $\vR$ and $\vx$ and equating these expressions to 0 provides the optimisation problems for: 1) the channel radii $\vR$, and 2) the location of the branching point $\vx$ (optimised in \app{sec:branching_point}). Optimisation of the channel radius is independent of $\vx$, because it is a decoupled problem \citep{Smink2023}, in which the power defined by \eq{eq:main branching power} attains a global minimum with respect to the radius of for each individual channel if each channel satisfies $\frac{\partial P}{\partial \vR}=\vec{0}$. Therefore, the optimisation condition for any radius $R_i$ only depends on the power contribution of the corresponding channel and $R_i$ does not depend on the lengths of the channels $L_i$.  

For minimizing the power consumption in the network, every single channel has to be optimised individually according to:
\begin{equation}
\pdv{P}{R_i}=\pdv{\Delta p_i}{R_i}Q_i+2\pi\alpha R_iL_i=0, \sp i=0,1,...,N.
\label{eq:main_deriv_P_R}
\end{equation}

The pressure drop $\Delta p$ is represented as a function of a flow rate $Q$ and a radius $R$ using the Darcy-Weisbach equation \citep{Weisbach1845}:
\begin{equation}
\Delta p = f\, \frac{\rho}{4\pi^2}\frac{Q^2}{R^5}L,
\label{eq:DW}
\end{equation}
Here, $\rho$ is the fluid density, $L$ is the length of the channel and $f$ the Darcy friction factor. The friction factor $f$ as function of the Reynolds number for the different fluid models is presented in \fig[b-d]{fig:figure_2}. The specific equations of $f$ for different flow regimes and fluid models are provided in \sec{sec:Flow_models}. In this study, a generalised Reynolds number is defined as \citep{Garcia1986}:

\be
Re\equiv \frac{8}{\pi^{2-n}}\left(\frac{n}{3n+1}\right)^n\frac{\rho}{\mu'}\frac{Q^{2-n}}{R^{4-3n}}.
\label{eq:main_def_Re}
\ee 
$f$ also depends on the Hedstr\"om number $He$ (commonly used for description of yield-stress fluids), the relative wall roughness $\delta \equiv \varepsilon/2R$ (where $\varepsilon$ is the absolute channel wall roughness) and the flow index $n$. 

The effective dynamic viscosity $\mu$ describes the resistance of a fluid against shear, and is defined as $\mu = \mu'|\dot{\gamma}|^{n-1}$, where $\dot{\gamma}$ is the shear rate, $\mu'$ is the flow consistency index and $n$ the flow index, where $0<n<1$ represents a shear-thinning, and $n>1$ represents a shear-thickening fluid as shown in \fig[a]{fig:figure_2}. In addition, a fluid may show yield-stress behaviour if $\tau_0\geq 0$ (e.g. Bingham and Herschel-Bulkley fluids, see \fig[a]{fig:figure_2}), where the fluid only shears once an effective local shear stress exceeds the yield stress $\tau_0$. The rheological behaviour for the shear rate of a Herschel-Bulkley fluid as function of the effective local shear stress is described as

\begin{equation}
 \dot{\gamma}(\tau) =
    \begin{cases}
      \text{sign}(\tau_{rz})\left( \frac{|\tau_{rz}|-\tau_0}{\mu'} \right)^{1/n} & \text{if $|\tau_{rz}|\geq\tau_0$},\\
       0 & \text{if $|\tau_{rz}|<\tau_0$}.\\
    \end{cases}       
    \label{eq:main_algemeen_model}
\end{equation}
where $\tau_{rz}$ is the local shear stress induced by the flow. \Fig[b-d]{fig:figure_2} also show the transition from laminar to turbulent flows at the critical Reynolds number \citep{Hanks1974}:
\be
Re_{crit} = \frac{6464n}{(1+3n)^2}\times (2+n)^{(2+n)/(1+n)}\times \frac{\psi^{2-n}}{(1-\phi)^{(n+2)/n}}.
\ee
where $\psi(\phi,n)$ and $\phi$ are defined in \sec{sec:laminar}. Although an increased relative roughness results in a decrease in $Re_{crit}$, this effect becomes only dominant around $\delta=0.1$ and larger \citep{Everts2022}. In order to ensure the validity of the used equations, the present study will limit the discussed optimisation results to $\delta\leq 0.1$.

%


The optimisation condition for the radius of every channel is obtained by substituting \eq{eq:DW} into \eq{eq:main_deriv_P_R}, providing:

\be
R^7=\frac{\rho Q^3}{8\pi^3\alpha}  B 
\label{eq:main_opt_cond_R}
\ee
where $B$ is defined as:
\be
B \equiv 5f -R\pdv{f}{R}.
\label{eq:main_def_B}
\ee
$B$ is a function of $f$ and $R$ and thus indirectly also a function of $Re$, $He$ $\delta$ and $n$.

The goal is to obtain an expression for the optimal channel radius as function of parameters that are independent of $R$. Therefore, dimensional analysis is carried out to find a dimensionless form of the channel radius as function of radius-independent dimensionless groups  (for more details, see \app{sec:subsec_nondim}). Scaling of the optimisation problem results in the following 5 dimensionless numbers:

\begin{multline}
\tilde{R} \equiv \frac{R}{\rho^{-\frac{1}{2}} \mu'^{\frac{3}{2(n+1)}} \alpha^{\frac{n-2}{2(n+1)}}},\sp
\tilde{Q} \equiv \frac{Q}{\rho^{-\frac{3}{2}} \mu'^{\frac{7}{2(n+1)}} \alpha^{\frac{3n-4}{2(n+1)}}},\quad
\tilde{\tau}_0 \equiv \frac{\tau_0}{\mu'^{\frac{1}{n+1}} \alpha^{\frac{n}{n+1}}},\quad\\
\tilde{\varepsilon} \equiv \frac{\varepsilon}{\rho^{-\frac{1}{2}} \mu'^{\frac{3}{2(n+1)}} \alpha^{\frac{n-2}{2(n+1)}}}, \text{ and } n
\label{eq:dimles_def}
\end{multline}

To align with literature, the Reynolds number will be used instead of $\tilde{R}$. This choice will simplify the later computations in \sec{sec:Flow_models} and enables direct characterisation of the flow in the channel. The Reynolds number is rewritten in dimensionless groups as follows:
\be
Re=a(n)\frac{\tilde{Q}^{2-n}}{\tilde{R}^{4-3n}},
\label{eq:main_def_Re_nondim}
\ee 
where $a(n)$ is defined as:
\be
a(n) \equiv \frac{8}{\pi^{2-n}}\left(\frac{n}{3n+1}\right)^n
\ee

The network is described by the dimensionless numbers $Re$, $\tilde{Q}$, $\tilde{\tau}_0$, $\tilde{\varepsilon}$ and $n$. As $\tilde{\tau}_0$, $\tilde{\varepsilon}$ and $n$ are fluid or system parameters that are constant in a network, $Re$ and $\tilde{Q}$ are the only flow-rate dependent dimensionless numbers, which is convenient for network optimisation. Described with the new dimensionless numbers, $f$ will now be a function of 5 dimensionless numbers instead of 4. Therefore, in the new domain, there will be a curve along which $f$ is constant, removing one degree of freedom (see \app{sec:4to5parameters}).

For given fluid, flow and system properties, an optimal channel radius can be calculated via the Reynolds number using the nondimensionalised optimisation condition (\eq{eq:main_opt_cond_R}):

\be
(8\pi^3)^{4-3n}\, a^7 \, \tilde{Q}^{2(n+1)}=  \tilde{B}^{4-3n}Re^7.
\label{eq:main_optimization_condition}
\ee
where $\tilde{B}$ is a function of $\tilde{f}$, which is a function of $Re$, $\tilde{\tau}_0$, $\tilde{Q}$, $\tilde{\varepsilon}$, and $n$. Hence, \eq{eq:main_def_B} can be rewritten as:

\be
\tilde{B} = 5\tilde{f}-R\pdv{\tilde{f}}{R} = 5\tilde{f}-R\,\pdv{\tilde{f}}{Re}\pdv{Re}{R} = \tilde{f}\left(5 +(4-3n)\frac{Re}{\tilde{f}} \,\pdv{\tilde{f}}{Re}\right)
\label{eq:main_def_B_dimless}
\ee
where differentiation of \eq{eq:main_def_Re} provides $\pdv{Re}{R} = -(4-3n)\frac{Re}{R}$. Therefore, when knowing $\tilde{\tau}_0$, $\tilde{Q}$, $\tilde{\varepsilon}$ and $n$, one can determine the optimal $Re$ for a channel. 

We conclude this section by introducing dimensionless groups that will be used in \sec{sec:Flow_models}. The Hedstr\"om number $He$ is defined \citep{Swamee_HB} and rewritten as follows:
\be
He \equiv \frac{4R^2\rho}{\mu'}\left(\frac{\tau_0}{\mu'}\right)^{\frac{2-n}{n}} =4\left(a(n)\frac{\tilde{Q}^{2-n}}{Re}\right)^{\frac{2}{4-3n}}\tilde{\tau}_0^{\frac{2-n}{n}}.
\label{eq:main_rewriting_He}
\ee
The relative roughness $\delta$ becomes in the new dimensionless groups:

\be
\delta\equiv \frac{\varepsilon}{2R} =\frac{1}{2}\, \tilde{\varepsilon} \left( \frac{Re}{a(n) \tilde{Q}^{2-n}}\right)^{\frac{1}{4-3n}}.
\ee

Characteristic for a yield-stress fluid is the formation of a plug in the centre of the channel, where the local shear stress $\tau_{rz}$ does not exceed the yield stress $\tau_0$. The associated plug radius $R_p$ is characterised by
\be
R_p \equiv \frac{2\tau_0}{\frac{\Delta p}{L}}.
\ee
Nondimensionalisation of the plug radius yields the dimensionless plug radius $\phi$, which often appears in descriptions for $f$ in the case of yield-stress fluids:

\be
\phi \equiv \frac{R_p}{R} = \frac{2\tau_0}{\frac{\Delta p}{L}R}.
\ee
By definition, $\phi$ can also be expressed in terms of the Reynolds and Hedstr\"om number and in the new dimensionless groups:

\be
\phi = \frac{64}{f\, Re}\left(\frac{2He}{Re}\left(\frac{n}{3n+1}\right)^2\right)^{\frac{n}{2-n}} = 8\pi^2 a(n)^{\frac{4}{4-3n}}\frac{\tilde{\tau}_0}{\tilde{f}\, Re}\left(\frac{\tilde{Q}^2}{Re^3}\right)^{\frac{n}{4-3n}}.
\label{eq:main_phi_inReHe}
\ee

Up to here, no assumptions are made with respect to the flow type. 

\section{Network optimisation for different flow regimes and fluid models}
\label{sec:Flow_models}

The following flow regimes and fluid models are discussed in subsections:
\begin{enumerate}
\item[\sec{sec:laminar}] Laminar flow of a Newtonian and non-Newtonian fluids.
\item[\sec{sec:turb_CW}] Turbulent flow of Newtonian fluids, rough and smooth channels.
\item[\sec{sec:turb_nN}] Turbulent flow of non-Newtonian fluids, smooth channel, for power-law and Herschel-Bulkley fluids.

\end{enumerate}

For each fluid model, we will obtain expressions for $\tilde{f}$, and thereafter for $\tilde{B}$, which via \eq{eq:main_optimization_condition} provides the optimal Reynolds number and therefore the optimal radius $R$ for each channel via \eqss{eq:main_def_Re}
{eq:dimles_def}. 
As many expressions for $\tilde{f}$ and $\tilde{B}$ are implicit, finding an optimal channel radius analytically is often impossible. Therefore, we also provide these results in the form of contour plots for the optimisation condition for $Re$ as a function of $\tilde{Q}$ and $\tilde{\tau}_0$, $\tilde{\varepsilon}$ or $n$.


\begin{figure}[h!]
\begin{center}
   \centering
   \includegraphics[width=1\textwidth]{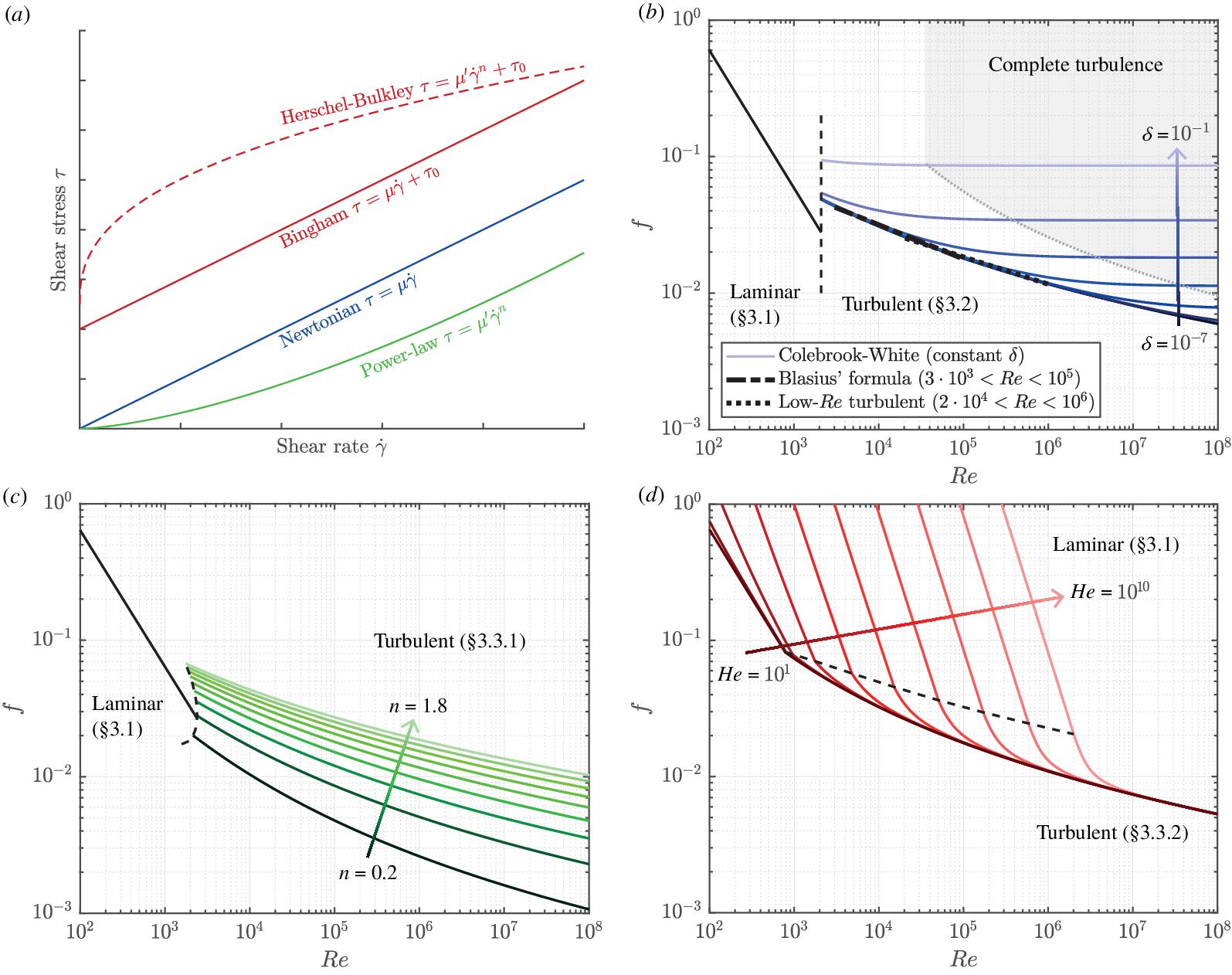} 
\caption{\scnr{a} Fluid models as analysed in this work. \scnr{b-d} Friction factor as function of the Reynolds number for different fluid models. The dashed black line indicates the transition from laminar to turbulent flow. \scnr{b} Newtonian fluid in rough channels (Moody diagram \citep{Moody1944}). The blue lines represent constant relative roughness $\delta$ from $10^{-7}, 10^{-6}, ... 10^{-1}$. \scnr{c} Power-law fluid in smooth channels. The green-scale lines represent values of the flow index $n$ from $0.2, 0.4, ... 1.8$. \scnr{d} Herschel-Bulkley fluid with $n=1.0$ in smooth channels. the red-scale lines represent constant values of the Hedstr\"om number from $10^{1}, 10^2, ..., 10^{10}$.}
\label{fig:figure_2}
\end{center}
\end{figure}

\begin{figure}[h!]
\begin{center}
   \centering
   \includegraphics[width=0.95\textwidth]{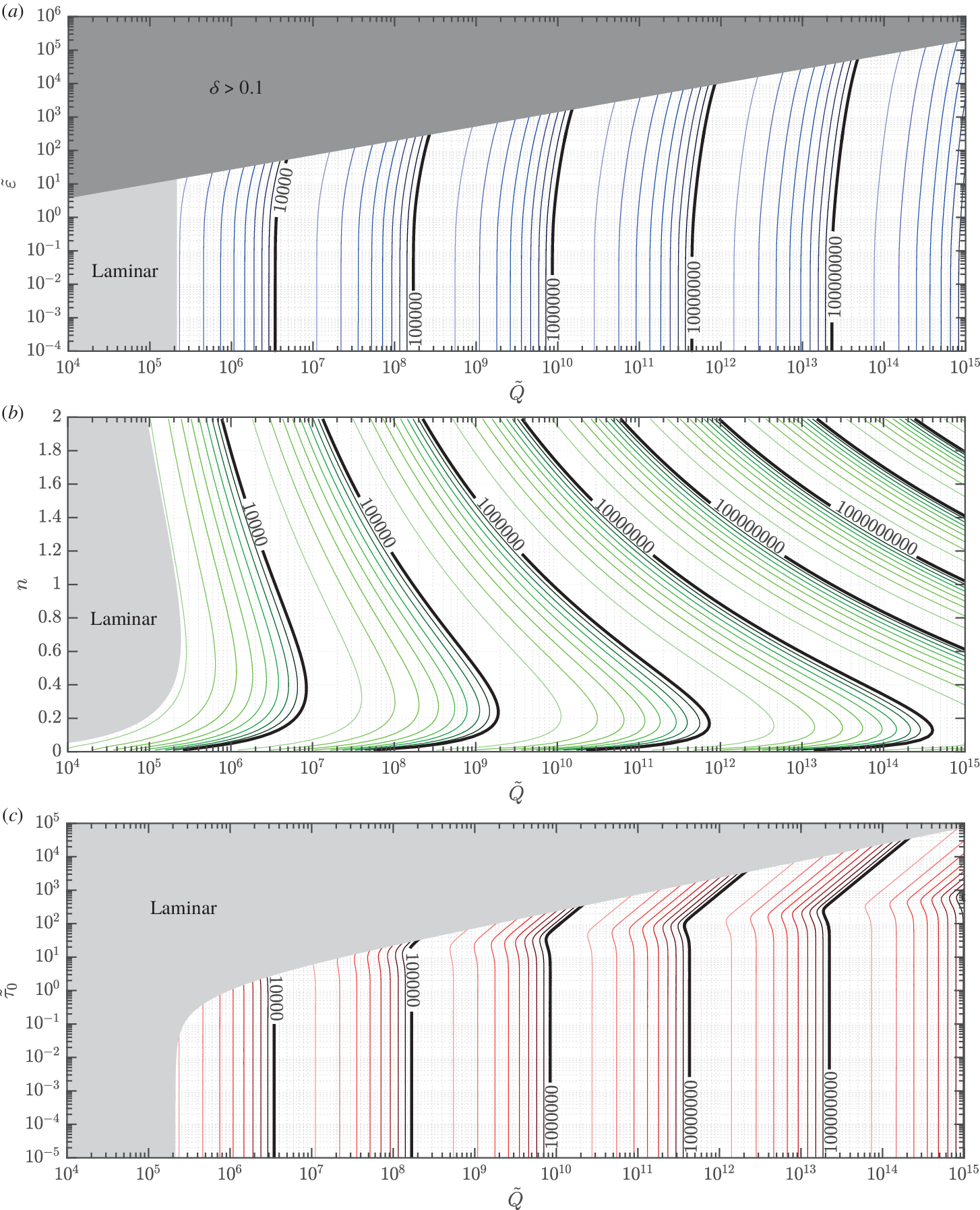} 
\caption{Optimal $Re$ for different fluid models. The colour-scale contour lines in represent the Reynolds numbers $2\times 10^x$, $3\times 10^x$, ... $9\times 10^x$ with decreasing brightness. \scnr{a} Contour plot of the optimal $Re$ as function of $\tilde{Q}$ and $\tilde{\varepsilon}$ for a Newtonian fluid in a rough-wall channel. \scnr{b} Contour plot of the optimal $Re$ as function of $\tilde{Q}$ and $n$ for a power-law fluid in a smooth-wall channel. \scnr{c} Contour plot of the optimal $Re$ as function of $\tilde{Q}$ and $\tilde{\varepsilon}$ for a Herschel-Bulkley fluid ($n=1$) in a smooth-wall channel.}
\label{fig:figure_3}
\end{center}
\end{figure}

\newpage
\subsection{Laminar flow of Newtonian and non-Newtonian fluids}
\label{sec:laminar}

For laminar flow through a circular channel results in the friction factor \citep{Smink2023}

\be
\tilde{f} =\frac{64}{Re \,\psi(\phi,n)^n},
\label{eq:f_laminar}
\ee
where $\psi$ is a dimensionless flow rate that ranges from 0 to 1. The friction factor is independent of the channel wall roughness, so it holds for arbitrary $\tilde{\varepsilon}$. For a Herschel-Bulkley fluid, $\psi$ is computed by integration of the velocity field over the channel's cross-section as \citep{Smink2023,Chilton1998,HerschelBulkley1926}:

\begin{equation}
  \psi = \frac{(1-\phi)^{(n+1)/n}}{(3n+1)^{-1}}
\times\left( \frac{(1-\phi)^2}{3n+1}+\frac{2\phi(1-\phi)}{2n+1}+\frac{\phi^2}{n+1}\right).
\label{eq:Turb_hb_psi_phi}
\end{equation}
For Bingham fluids ($n=1$), $\psi$ reduces to \citep{Smink2023,Reiner1926}:

\begin{equation}
  \psi = 1-\frac{4}{3}\phi+\frac{1}{3}\phi^4  , 
  \label{eq:Turb_psi_bingham}
\end{equation}
and for the case of $\tau_0=0$ (Newtonian and power-law fluids, $\phi=0$), $\psi$ further reduces to 1.

Calculation of the optimisation condition results in the following expression for $\tilde{B}$:

\be
\tilde{B} = \tilde{f}J(\phi,n),
\ee
where the function $J(\phi,n)$ is defined as

\be
J\equiv 1+\frac{3n}{1-\frac{n\phi}{\psi}\pdv{\psi}{\phi}}.
\label{eq:Turb_def_J}
\ee

For a Herschel-Bulkley fluid, substitution of $\psi$ from \eq{eq:Turb_hb_psi_phi} into \eq{eq:Turb_def_J} results in the following expression for $J$:
\be
J = \frac{3n+1}{\frac{6n^3}{(2n+1)(n+1)}\phi^3+\frac{6n^2}{(2n+1)(n+1)}\phi^2+\frac{3n}{2n+1}\phi +1}.
\label{eq:Turb_J_HB}
\ee
For Bingham fluids ($n=1$), the expression for $J$ reduces to 
\be
J = \frac{4}{\phi^3+\phi^2+\phi +1}.
\label{eq:Turb_J_B}
\ee
and for non-yield fluids ($\phi=0$), it reduces to $J=3n+1$. Finally, for the yield limit ($\phi=1$), one obtains $J=1$.

Substitution of $\tilde{B}$ into \eq{eq:main_optimization_condition} reduces the optimisation condition to

\be
Re^{3(n+1)}=\left(\frac{\pi^3}{8}\right)^{4-3n}\, a^7\, \tilde{Q}^{2(n+1)} \left(\frac{\psi(\phi,n)^n}{J(\phi,n)}\right)^{4-3n}.
\label{eq:laminar_opt_Re}
\ee

Together with \eq{eq:main_phi_inReHe} and \eq{eq:f_laminar}, $Re$ is then explicitly solved for given $\tilde{Q}$, $\tilde{\tau}_0$ and $n$, providing the optimal radius via \eqss{eq:main_def_Re}{eq:dimles_def}. \Fig[a-c]{fig:laminar_all} in \app{app:extra_figures_HB} present a contour plot for the optimal $Re$ as function of $\tilde{Q}$ and $\tilde{\tau}_0$ for different values of $n$. Note that the number of parameters can be reduced even further, by combining $\tilde{Q}$ and $Re$, to a parameter which is only a function of given $\tilde{\tau}_0$ and $n$. Then the optimisation condition becomes \citep{Smink2023}:

\be
\frac{\tilde{R}^3}{\tilde{Q}} =\left( a^3\frac{\tilde{Q}^2}{Re^3}\right)^{\frac{1}{4-3n}}= \frac{1}{\pi}\left(\left(\frac{3n+1}{n}\right)^n \frac{J(\phi,n)}{\psi(\phi,n)^n}\right)^{1/(n+1)},
\label{eq:laminar_optimization}
\ee
where the right-hand side is constant for an optimised network. Plotting $\frac{\tilde{R}^3}{\tilde{Q}}$ as function of $\tilde{\tau}_0$ and $n$ gives the contour plot in \fig[d]{fig:laminar_all} in \app{app:extra_figures_HB}. This figure can be made for laminar flows because $\tilde{R}^3/\tilde{Q}$ is constant over the entire network, which is highly useful for design of optimised branched fluidic networks, as only one optimisation condition has to be calculated.

\subsection{Turbulent flow of Newtonian fluid, rough and smooth channel}
\label{sec:turb_CW}

We now demonstrate that the proposed optimisation method is universal if the friction factor is known, by optimizing networks for diverse turbulent flow regimes. For turbulent flow of a Newtonian fluid, the Colebrook-White equation \citep{Colebrook1939} describes the friction factor for both low- and high-turbulent flow ($Re = [Re_{crit},\infty\rangle$) in hydraulically smooth and rough pipes using the following implicit equation:

\be
f = \left\{-2.0\log_{10}\left(\frac{\delta}{3.7}+\frac{2.51}{Re\sqrt{f}}\right)\right\}^{-2}.
\label{eq:Colebrook_White}
\ee

\Eq{eq:Colebrook_White} underlies the well-known Moody diagram \citep{Moody1944} as shown in \fig[b]{fig:figure_2}. Rewriting \eq{eq:Colebrook_White} in terms of the dimensionless numbers gives:

\be
\tilde{f} = \left\{-2.0\log_{10}\left(\frac{1}{3.7}\,\frac{\pi}{4}\frac{\tilde{\varepsilon}Re}{\tilde{Q}}+\frac{2.51}{Re\sqrt{\tilde{f}}}\right)\right\}^{-2}.
\label{eq:Colebrook_White_nondim}
\ee
Differentiation to $Re$ gives:

\be
\pdv{\tilde{f}}{Re} = -\frac{\tilde{f}}{Re}\frac{2\left(\frac{1}{3.7}\,\frac{\pi}{4}\frac{\tilde{\varepsilon}Re}{\tilde{Q}}-\frac{2.51}{Re\sqrt{\tilde{f}}}\right)}{\left(\frac{1}{3.7}\,\frac{\pi}{4}\frac{\tilde{\varepsilon}Re}{\tilde{Q}}+\frac{2.51}{Re\sqrt{\tilde{f}}}\right)\ln\left(\frac{1}{3.7}\,\frac{\pi}{4}\frac{\tilde{\varepsilon}Re}{\tilde{Q}}+\frac{2.51}{Re\sqrt{\tilde{f}}}\right)-\frac{2.51}{Re\sqrt{\tilde{f}}}}
\ee
resulting in

\be
\tilde{B} = \tilde{f}\left(5-\frac{2\left(\frac{1}{3.7}\,\frac{\pi}{4}\frac{\tilde{\varepsilon}Re}{\tilde{Q}}-\frac{2.51}{Re\sqrt{\tilde{f}}}\right)}{\left(\frac{1}{3.7}\,\frac{\pi}{4}\frac{\tilde{\varepsilon}Re}{\tilde{Q}}+\frac{2.51}{Re\sqrt{\tilde{f}}}\right)\ln\left(\frac{1}{3.7}\,\frac{\pi}{4}\frac{\tilde{\varepsilon}Re}{\tilde{Q}}+\frac{2.51}{Re\sqrt{\tilde{f}}}\right)-\frac{2.51}{Re\sqrt{\tilde{f}}}}\right).
\ee
Substituting $\tilde{B}$ into \eq{eq:main_optimization_condition} enables implicit solving $Re$ for given $\tilde{Q}$. Contour plots of the optimal $Re$ as function $\tilde{Q}$ and $\tilde{\varepsilon}$ are presented in \fig[a]{fig:figure_3}. Limit cases of the Colebrook-White equation include complete turbulence (for which Von Kármán's formula applies), turbulence in smooth channels, and low-Reynolds-number turbulence in smooth channels (e.g. Blasius' formula). Expressions of the optimal channel radius for these cases are derived in \app{app:specific_Newt_turb}. 

\subsection{Turbulent flow of non-Newtonian fluids, smooth channel}
\label{sec:turb_nN}
\subsubsection{Power-law fluid}
\label{sec:turb_powerlaw}
Turbulent pipe flow of non-Newtonian power-law fluids at low Reynolds numbers in a smooth circular channel ($\varepsilon\rightarrow 0$) was analysed by amongst others Dodge and Metzner \citep{Dodge1959}. They modified the Von Kármán equation for turbulent Newtonian pipe flow, resulting in an implicit relation for the Darcy friction factor $f$:
\be
\frac{2}{\sqrt{f}}=\frac{4}{n^{0.75}}\log_{10}\left(Re\left(\frac{f}{4}\right)^{1-n/2}\right)-\frac{0.4}{n^{1.2}}
\label{eq:DodgeMetzner}
\ee
$f$ is shown as a function of $Re$ and $n$ in \fig[c]{fig:figure_2}.

Implicit differentiation of $\tilde{f}$ to $Re$ gives

\be
\pdv{\tilde{f}}{Re} = -\frac{\tilde{f}}{Re}\left( \frac{n^{0.75}\ln 10}{4\sqrt{\tilde{f}}}+1-\frac{n}{2} \right)^{-1}
\ee
resulting in

\be
\tilde{B} = \tilde{f}\left(5+(3n-4)\left( \frac{n^{0.75}\ln 10}{4\sqrt{\tilde{f}}}+1-\frac{n}{2} \right)^{-1}\right)
\ee

Here, $\tilde{B}$ is still a function of $\tilde{f}$, which is in turn an implicit function of $Re$ governed by \eq{eq:DodgeMetzner}. Therefore, $\tilde{B}$ is governed by $Re$ and the flow index $n$. Consequently, with this expression for $\tilde{B}$, the optimal value of $Re$ in a channel can be calculated as function of $\tilde{Q}$ and $n$. A contour plot for the optimal Reynolds number in a channel is given in \fig[b]{fig:figure_3}.

%

\subsubsection{Herschel-Bulkley fluid}
\label{sec:turb_HB}

For a turbulent flow of a Herschel-Bulkley fluid in a smooth circular channel ($\varepsilon \rightarrow 0$), Torrance \citep{Garcia1986} developed a relationship for the friction factor. This relation for the Darcy friction factor is given by:

\be
\frac{2}{\sqrt{\tilde{f}}}=0.45-\frac{2.75}{n}+\frac{1.97}{n}\ln(1-\phi) +\frac{1.97}{n}\ln\left(Re\left(\frac{3n+1}{4n}\right)^n \left(\frac{\tilde{f}}{4}\right)^{1-\frac{n}{2}}\right).
\label{eq:Torrance}
\ee
Here, $\phi$ as given in \eq{eq:main_phi_inReHe} is used in the calculations. \Fig[d]{fig:figure_2} shows $f$ as function of $Re$ for $n=1.0$ and a range of values of $\tilde{\tau}_0$. $f$-$Re$ plots for other values of $n$ are presented in \fig{fig:f_Re_plots_HB} in \app{app:extra_figures_HB}.

Implicit differentiation of $\tilde{f}$ to $Re$ leads to

\be
\frac{\partial \tilde{f}}{\partial Re} = -\frac{\tilde{f}}{Re}\frac{1+\frac{\phi}{1-\phi}\frac{4}{4-3n}}{\frac{n}{1.97}\frac{1}{\sqrt{\tilde{f}}}+\frac{\phi}{1-\phi}+1-\frac{n}{2}},
\ee
giving an expression for $\tilde{B}$:

\begin{equation}
\tilde{B} = \tilde{f}\left(5-\frac{4-3n+\frac{4\phi}{1-\phi}}{\frac{n}{1.97}\frac{1}{\sqrt{\tilde{f}}}+\frac{\phi}{1-\phi}+1-\frac{n}{2}}\right).
\end{equation}

Contour plots for the optimal Reynolds number in a channel as function of $\tilde{\tau}_0$ and $\tilde{Q}$ are given for $n=1.0$ in \fig[c]{fig:figure_3} and the results for $n=0.5$ and $n=1.5$ are provided in \fig{fig:contour_plots_HB} in \app{app:extra_figures_HB}. An additional fluid model covering laminar and turbulent flow of a Bingham fluid is presented in \app{sec:Darby}.

\section{Scaling of the channel radii}
\label{sec:Synthesis}

In the literature, many attempts have been made to find the proportionality between the flow rate $Q$ and the channel radius $R$ for optimised networks, in the form of \eq{eq:def_x} \citep{Sherman1981,Uylings1977,Woldenberg1986,Williams2007, Kou_etal2014}. In the optimisation method of the present study, only $Re$ and $\tilde{Q}$ contain parameters involving $R$ and $Q$, while $\tilde{\tau}_0$, $\tilde{\varepsilon}$ and $n$ are independent of $R$ and $Q$. Therefore, when knowing the proportionality between $\tilde{Q}$ and $Re$, $x$ can be calculated analytically. However, this only holds if $\tilde{B} \propto Re^{c_1}\tilde{Q}^{c_2}$, which is only true for relatively simple descriptions of $f$ applicable to laminar flows or limit cases of the turbulent regime (e.g. Blasius' formula, \app{app:Blasius}). Therefore, in the following, $x$ will be calculated by locally approximating $\tilde{B}$ by a power function via

\be
x = \frac{R}{Q}\frac{\partial Q}{\partial R}= \frac{(4-3n)\frac{Re}{\tilde{Q}}\pdv{\tilde{Q}}{Re}}{(2-n)\frac{Re}{\tilde{Q}}\pdv{\tilde{Q}}{Re}-1}. 
\label{eq:main_x_def}
\ee

\begin{figure}[h!]
\begin{center}
   \centering
   \includegraphics[width=1.0\textwidth]{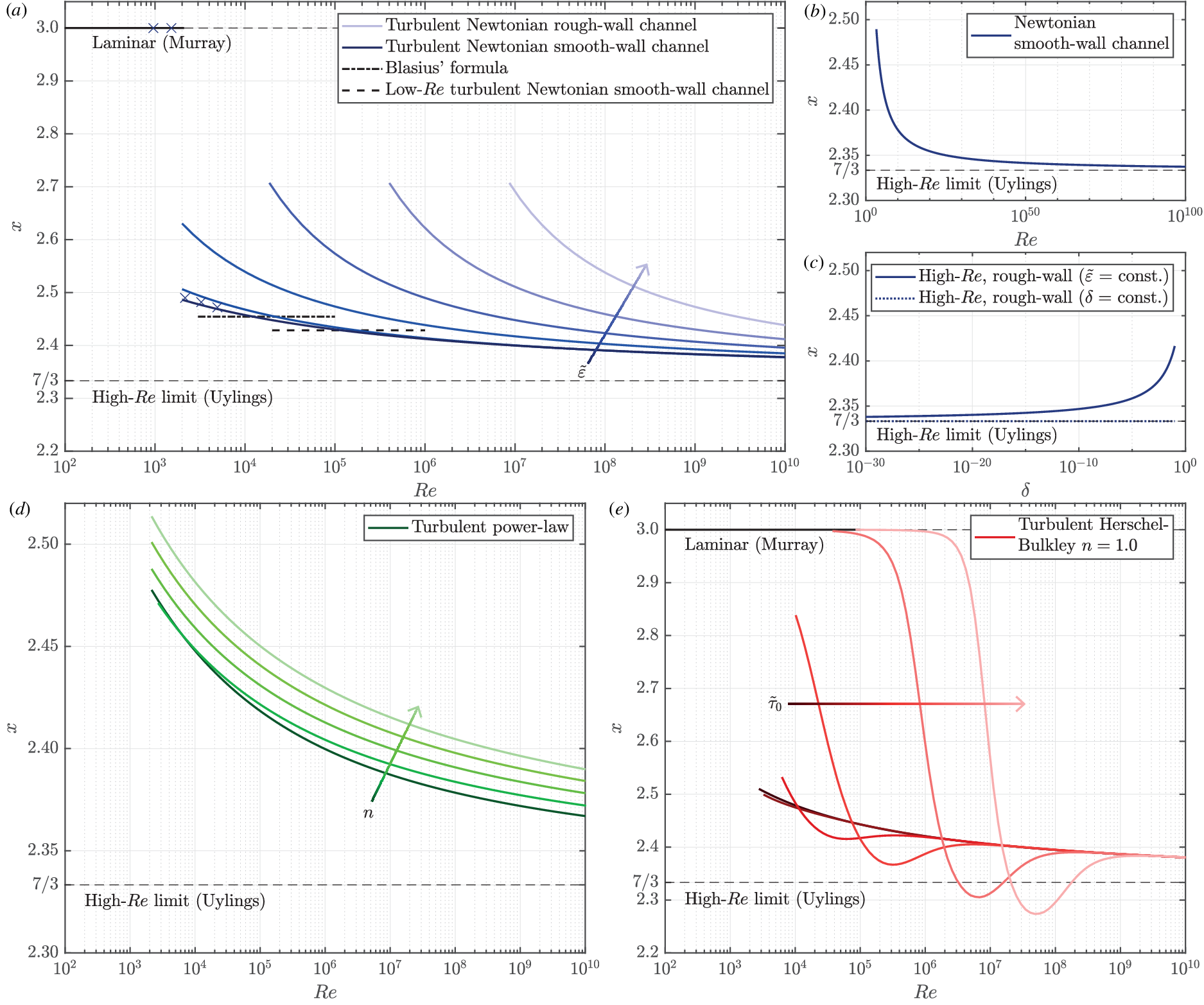} 
\caption{Plot of $x$ (\eq{eq:main_x_def}) as function of $Re$ for the different fluid models as discussed in \sec{sec:Flow_models}. The dashed lines for $x=3$ and $x=7/3$ show the expected limit cases for laminar flow and high-turbulent flow, respectively \citep{Uylings1977}. \scnr{a} Turbulent flow of Newtonian fluids, described by Colebrook-White ($\tilde{\varepsilon} \in [10^{-5},10^{1},10^{2},10^{3},10^{4},10^{5}]$), Blasius' formula (\eq{eq: Blasius formula}) and an empirical relation (\eq{eq:empirical_approx_Newt}). The blue $\times$-symbols correspond with the optimised channels in \fig[e]{fig:overview}, showing that the values for $x$ change significantly for that network. \scnr{b} Plot of $x$ as function of $\Re$ for turbulent flow of a Newtonian fluid in a hydraulically smooth channel, described by \eq{eq:Colebrook_White} for the limit of $\tilde{\varepsilon}=0$. For optimised networks, high-$Re$ turbulent flow in smooth channels will for realistic $Re$ never reach $x=7/3$. \scnr{c} Plot of $x$ as function of $\delta = \varepsilon/2R$ for high-$Re$ turbulent flow of a Newtonian fluid in a rough channel, described by \eq{eq:VonKarman}. For optimised networks, high-$Re$ turbulent flow in channels with finite roughness will never reach $x=7/3$. \scnr{d} Turbulent flow of a power-law fluid (Metzner \& Dodge) ($n\in [0.2,0.4,1.0,1.4,1.8]$). \scnr{e} Turbulent flow of a Herschel-Bulkley fluid (Torrance) ($\tilde{\tau}_0 \in [0.1,0.25,1,2.5,10,25]$ with $n= 1.0$). Plots for $n=0.5$ and $n=1.5$ are presented in \fig{fig:X_HB} in \app{app:extra_figures_HB}.}
\label{fig:x_figure}
\end{center}
\end{figure}

By calculating the fluid-model specific $\frac{Re}{\tilde{Q}}\pdv{\tilde{Q}}{Re}$, one obtains $x$ in the relevant ($Re,\tilde{\tau}_0,\tilde{\varepsilon},n$)-space, as shown in \fig{fig:x_figure} for the different fluid models (the underlying equations are shown in \app{app:deriv_x}). Known limit cases are shown in black; all coloured results are new. For laminar flow of all treated fluid models, $x$ becomes 3, as expected \citep{Murray1926a,Smink2023} For networks that span ranges of $Re$ for which $x$ is (almost) constant, the value of $x$ is readily used to scale all channels in the network if one optimal diameter is known, using $Q\propto R^x$.  

For a Newtonian fluid (\fig[a]{fig:x_figure}), at the transition from laminar to turbulent flow, $x$ drops quickly from 3 to around 2.5 for hydraulically smooth channels.  We find that the low-$Re$ approximations (Blasius, $x=27/11$, and an empirical relation, $x=17/7$ \citep{Kou_etal2014}) correspond to the average value of $x$ over their applicable $Re$-range of the Colebrook-White equation for a smooth channel. Although these approximations and the resulting values of $x$ are valuable for networks that span a limited range of the Reynolds number, the network shown in \fig[e]{fig:overview} represents a case where $x$ is not applicable for determination of the optimal channel radii. Here, the optimal network partly corresponds to laminar flow and partly to turbulent flow. As these regimes result in strongly different values of $x$, as shown by the markers ($\times$) in \fig[a]{fig:x_figure}, the so-called Murray's law \citep{Sherman1981,Stephenson_and_Lockerby2016} ($R_0^x = \sum_i R_i^x$, for one branching point) will not hold. Therefore, the optimal radius of each channel must be calculated individually.


\Fig[a]{fig:x_figure} also shows that $x$ increases for increasing wall roughness. The lines are clipped for $\delta = 0.1$, corresponding to $x \approx 2.71$.  Analysing the Colebrook-White equation for the smooth channel limit ($\tilde{\varepsilon}=0$) shows that for extremely large $Re$, $x$ will indeed approach $x=7/3$, as shown in \fig[b]{fig:x_figure}. The high-$Re$ limit for nonzero roughness however (\fig[c]{fig:x_figure}, $\tilde{\varepsilon}= \text{const.}$) shows that $x=7/3$ will never be reached if a finite value of $\delta$ is included in the optimisation process.

When using the Colebrook-White equation, one has to make an assumption about the wall roughness. \Fig[a]{fig:x_figure} shows the results for a fixed dimensionless absolute roughness $\tilde{\varepsilon}$. One can also choose to fix the relative wall roughness $\delta$, which has often been an implicit assumption in previous works \citep{Uylings1977,Bejan2000, Stephenson_and_Lockerby2016,Kou_etal2014}. Then $\delta$ is excluded from the optimisation procedure. Only in this case, it holds for complete turbulence in rough-wall channels that $x=7/3$ (\fig[c]{fig:x_figure}, $\delta = \text{const.}$). The assumption of a fixed $\delta$ requires that the absolute roughness $\varepsilon$ is proportional to the optimised channel radius for every channel in the network. However, many networks consist of a given wall material with uniform absolute roughness $\varepsilon$. Even when there is a changing absolute roughness in the network, the choice for the use of $\tilde{\varepsilon}$ is preferred and then the graphical approach (\fig[a]{fig:figure_3}) can be used for optimisation of the channel radii.

\Fig[d]{fig:x_figure} shows $x$ for turbulent flow of a power-law fluid in smooth channels. The behaviour of $x$ is similar to that of a Newtonian fluid in a smooth channel, but increases slightly with increasing $n$. \Fig[e]{fig:x_figure} shows the results for a Herschel-Bulkley fluid ($n=1.0$), which appears to have a complete transition from $x=3$ to $x\rightarrow 7/3$, with a remarkable bump in the transition, which is related to the sharp corner around $10^1<\tilde{\tau}_0<10^2$ as observed in the optimisation plot (\fig[c]{fig:figure_3}). The transition occurs at larger $Re$ for increasing $\tilde{\tau}_0$, and for large $Re$, all lines converge. Results for $n=0.5$ and $n=1.5$ are presented in \fig{fig:X_HB} in \app{app:extra_figures_HB}.

The transitional behaviour of $x$ as presented in \fig{fig:x_figure} stresses its importance for network optimisation when considering turbulent flow. For domains of $Re$ where $x$ is approximately constant, \fig{fig:x_figure} enables making a quick estimation of the optimal channel radii within a network. If one channel within a network has been optimised, then the other channels can be optimised using $Q\propto R^x$. For domains of $Re$ where $x$ clearly non-constant, the full computations of the equations or the use of the graphical approach of \fig{fig:figure_3} is required.

\section{Synthesis: Design of an optimised fluidic network}
\label{sec:designprocedure}

The framework described in \secsss{sec:main_grote_sectie_optimalisatie}{sec:Flow_models}{sec:Synthesis} leads to the following approach for optimisation of branched fluidic networks (such as shown in \fig[e,f]{fig:overview}):

\begin{enumerate}
\item Determine the fluid model and the cost factor $\alpha$, and calculate $\tilde{\tau}_0$, $\tilde{\varepsilon}$ and $n$. Calculate $\tilde{Q}$ for all the to be optimised channels (\sec{sec:main_grote_sectie_optimalisatie}).
\item Determine the optimal radius $R$ in one channel.
\begin{enumerate}
\item First, determine the optimal $Re$ by one of the following options:
\begin{itemize}
\item[-] Read $Re$ from an optimisation contour plot (e.g. \fig[a]{fig:figure_3}) for given $\tilde{Q}$.
\item[-] Calculate $\tilde{B}(\tilde{f},Re,\tilde{Q},\tilde{\tau}_0,\tilde{\varepsilon},n)$ via \eq{eq:main_def_B_dimless} and then solve \eq{eq:main_optimization_condition} for $Re$ for given $\tilde{Q}$.

\end{itemize}
\item Subsequently, calculate $R$ from the calculated $Re$ by using \eq{eq:main_def_Re}.
\end{enumerate}

\item Obtain all channel radii:
\begin{itemize}
\item[-] By repeating step 2 for each channel, or
\item[-] by obtaining $x$ from \fig{fig:x_figure} and calculating all channel radii using $Q\propto R^x$. This option only applies if $x$ is (almost) constant over the applicable range of $Re$.
\end{itemize}
\item Calculate the optimal location of the branching points (\app{sec:branching_point}).
\end{enumerate}

\section{Conclusion}
\label{sec:Turb_conclusion}

A unifying design procedure for optimisation of branched fluidic networks was presented. The optimal channel radius is obtained by minimizing the power consumption for maintaining the flow and minimizing the volume simultaneously, which is carried out for all channel segments individually. The approach is applicable to every flow configuration for which the Darcy friction factor is available.

In this work, network optimisation was carried out for the parameter space spun by the Reynolds number (laminar and turbulent flows), the wall roughness and for different rheologies. This parameter space is widely applicable to e.g. heat exchangers or (intertial) microfluidics, but was hardly covered in the literature as shown in \fig[a,b]{fig:overview}. First, the entire optimisation problem is non-dimensionalised in terms of $Re$, $\tilde{Q}$, $\tilde{\tau}_0$, $\tilde{\varepsilon}$ and $n$. For every channel, the optimal Reynolds number and the optimal channel radius are obtained as a function of the dimensionless flow rate $\tilde{Q}$. To minimise the required mathematical analysis, graphs are presented that readily provide the optimal Reynolds number for every fluid model. 

Subsequently, the value of $x$ in the proportionality $Q\propto R^x$ was evaluated for the full parameter space of the Reynolds number and channel roughness for different fluid rheologies.  The well-known limit values $x=3$ for laminar flow and $x=7/3$ for high-$Re$ turbulence were recovered \citep{Uylings1977,Bejan2000, Stephenson_and_Lockerby2016,Kou_etal2014}. However, as appears that the latter value is hardly reached in practice, the intermediate values of $x$ were calculated and plotted. Finally, a design procedure of optimised fluidic networks that incorporates all aforementioned elements was presented.

Our unifying approach for optimisation of fluidic networks can be extended to other configurations with known friction factors, including non-circular cross-sections, bended pipes, bubbly flows, porous walls \citep{Miguel2018}, porous media \citep{Macdonald1991} or slip flow \citep{Spiga1998}. As such, it opens an enormous parameter space of realistically-encountered configurations in nature and engineering.

\subsection*{Acknowledgements}
E.A. Bramer, K. Jain, and T.G. Vlogman are acknowledged for the helpful discussions.

\printbibliography[heading=bibintoc]


\appendix
\newpage

\section*{\apptitle}
\section{Details onto optimisation of a fluidic branching}
\label{sec:Turb_grote_sectie_optimalisatie}

\subsection{Optimal branching point location}
\label{sec:branching_point}

To minimise the power in a network, the location of the branching point $\vx$ must be chosen. If the channel radii are optimised according to \eq{eq:main_optimization_condition}, this is the same as minimizing the total volume within the channels. Differentiation of the total power $P$ (\eq{eq:main branching power}) to $\vx$ results in:
\be
\nabla_{\vx}{P}=\sum_{i=0}^N\left(\left|\dv{p}{z}\right|_i Q_i + \alpha\pi R_i^2\right) \vnabla L_i=\textbf{0}, \sp i=0,1,...,N,
\label{eq:Turb_nabla_P}
\ee
where we define
\be
\ve_{i,*}\equiv\left(\vnabla L_i\right)_*=\frac{\vx_*-\vx_i}{|\vx_*-\vx_i|}.
\ee
Here, the subscript $i$ indicates the index of the channel, and the subscript $*$ indicates an optimised parameter.

Substituting the optimisation condition for the channel radii (\eq{eq:main_optimization_condition}) results in the optimisation condition of the branching point $\vx=\vx_*$ (for a full derivation, see \citet{Smink2023}):
\be
\sum_{i=0}^N \left(R_{i,*}^2 \frac{2f+B}{B} \ve_{i,*}\right)=\textbf{0}, \sp i=0,1,...,N.
\label{eq:branching_point}
\ee
Alternatively, we can rewrite \eq{eq:branching_point} in terms of dimensionless numbers:
\be
\sum_{i=0}^N \left(\frac{\tilde{Q}^{2-n}}{Re}\right)^{\frac{2}{4-3n}} \frac{2\tilde{f}+\tilde{B}}{\tilde{B}} \ve_{i,*}=\textbf{0}, \sp i=0,1,...,N.
\label{eq:branching_point1}
\ee
\Eq{eq:branching_point1} requires the optimal Reynolds number and corresponding $\tilde{f}$, $\tilde{B}$ and $\tilde{Q}$ for each channel for computation of the optimal branching point location. It is an implicit equation, such that the coordinates of $\vx_*$ are solved by simple numerical methods, providing the lengths of the channels $L_i$. The resulting $\vx_*$ determines a network that is optimised both with respect to the channel radii (\eq{eq:main_optimization_condition}) and the channel lengths (\eq{eq:branching_point1}).

\subsection{Governing dimensionless numbers}
\label{sec:subsec_nondim}

Scaling of the problem for calculation of the optimisation condition for the channel radii is investigated using the Buckingham-Pi theorem. The goal is to find a nondimensionalisation of the radius and the dimensionless parameters where it depends on. In this problem, the following 8 independent parameters are relevant: The channel radius $R$ in m, the flow rate $Q$ in m$^3$ s$^{-1}$, the fluid density $\rho$ in kg m$^{-3}$, the viscosity index $\mu'$ in Pa s$^n$, the cost factor $\alpha$ in W m$^{-3}$, the yield stress $\tau_0$ in Pa, the channel absolute roughness $\varepsilon$ in m and the flow index $n$. In these parameters, 3 independent physical dimensions are present. We choose $\rho$, $\mu'$, and $\alpha$ to scale this problem, which are suitable for the scaling. We will end up with 5 dimensionless parameters, where $n$ is one of them.

In mathematical terms, the problem is then:
\begin{equation}
R = g(\rho,\mu',\alpha,Q,\tau_0,\varepsilon,n),
\end{equation}
where scaling of the problem results in:
\begin{equation}
\tilde{R} = \tilde{g}(\tilde{Q},\tilde{\tau_0},\tilde{\varepsilon},n)
\end{equation}
with 

\begin{multline}
\tilde{R} \equiv \frac{R}{\rho^{-\frac{1}{2}} \mu'^{\frac{3}{2(n+1)}} \alpha^{\frac{n-2}{2(n+1)}}},\sp
\tilde{Q} \equiv \frac{Q}{\rho^{-\frac{3}{2}} \mu'^{\frac{7}{2(n+1)}} \alpha^{\frac{3n-4}{2(n+1)}}},\quad\\
\tilde{\tau}_0 \equiv \frac{\tau_0}{\mu'^{\frac{1}{n+1}} \alpha^{\frac{n}{n+1}}},\text{ and }
\tilde{\varepsilon} \equiv \frac{\varepsilon}{\rho^{-\frac{1}{2}} \mu'^{\frac{3}{2(n+1)}} \alpha^{\frac{n-2}{2(n+1)}}}
\end{multline}

For convenience, the Reynolds number will be used instead of $\tilde{R}$. This choice simplifies the calculations and provides direct characterisation of the flow in the channels. Expressed in dimensionless groups, the Reynolds number (\eq{eq:main_def_Re}) becomes:
\be
Re= a(n)\frac{\tilde{Q}^{2-n}}{\tilde{R}^{4-3n}}
\ee
where $a(n)$ is defined as:
\be
a(n) \equiv \frac{8}{\pi^{2-n}}\left(\frac{n}{3n+1}\right)^n
\ee

In the end, a network can be described by the dimensionless numbers $Re$, $\tilde{Q}$, $\tilde{\tau}_0$, $\tilde{\varepsilon}$ and $n$. All parameters except for $R$ and $Q$ are fluid or system parameters, which are constant in a network. In a branched fluidic network, $Q$ can change from channel to channel. Therefore, for design procedure purposes, it is convenient that $Re$ and $\tilde{Q}$ are the only flow-rate dependent dimensionless numbers. 

Therefore, the optimisation problem is reduced to the dimensionless dependency
\begin{equation}
Re = \tilde{g}_1(\tilde{Q},\tilde{\tau}_0,\tilde{\varepsilon},n).
\end{equation}

For $n=1$, the dimensionless numbers reduce to:

\begin{equation}
Re  = \frac{2}{\pi}\frac{\rho Q}{\mu' R}, \quad \tilde{Q} = \frac{\alpha^{\frac{1}{4}}\rho^{\frac{3}{2}} Q}{\mu'^{\frac{7}{4}}},\quad \tilde{\tau}_0 = \frac{\tau_0}{\left(\mu'\alpha\right)^{\frac{1}{2}}}, \text{ and } \tilde{\varepsilon} = \frac{\rho^{\frac{1}{2}}\alpha^{\frac{1}{4}}\varepsilon}{\mu'^{\frac{3}{4}}}
\end{equation}

For given fluid, flow and system properties, an optimal channel radius can be calculated from the obtained optimal Reynolds number.

\subsection{Mapping from 4 to 5 variables for $f$}
\label{sec:4to5parameters}

The friction factor $f$ commonly depends on the Reynolds number $Re$, the Hedstr\"om number $He$, the relative wall roughness $\delta$ and the flow index $n$. These are 4 independent dimensionless groups. In the optimisation procedure, the following 5 dimensionless numbers are used (for details, see \app{sec:subsec_nondim}): $Re$, $\tilde{Q}$, $\tilde{\tau}_0$, $\tilde{\varepsilon}$ and $n$. Because the friction factor will now be described by 5 instead of 4 dimensionless numbers, a curve in the 5-variables space exists along which $f$ is constant. The direction of that curve is locally described by the vector $\vec{\beta} = (\beta_1, \beta_2,\beta_3, \beta_4, \beta_5)^T$. Therefore, the inner product of this vector with the gradient of the original variables $\vec{F} = (Re,He, \delta,n)^T$ to the new variables $\tilde{\vec{F}} = (Re,\tilde{Q},\tilde{\tau}_0,\tilde{\varepsilon},n)^T$ must be zero. In that case, $Re$, $He$, $\delta$ and $n$ are constant along the direction of $\vec{\beta}$. This results in the following 4 equations:

\be
\sum_{j=1}^5\pdv{F_i}{\tilde{F}_j}\beta_j =0, \sp\sp\sp i = 1, 2, 3, 4,
\label{eq:mapping}
\ee
which are sufficient to define the direction of the vector $\vec{\beta}$ while choosing its arbitrary length by conveniently choosing $\beta_2 =1$. Solving \eq{eq:mapping} results in $\beta_1 = 0$, $\beta_3 = -\frac{2n}{4-3n}\frac{\tilde{\tau}_0}{\tilde{Q}}\beta_2$, $\beta_4 = \frac{2-n}{4-3n}\frac{\tilde{\varepsilon}}{\tilde{Q}}\beta_2$, and $\beta_5 = 0$.

$f$ will be restricted in the new parameter space by:
\be
\sum_{j=1}^5\pdv{f}{\tilde{F}_j}\beta_j =0.
\ee
Substituting the obtained coefficients $\vec{\beta}$ results in the following expression:

\be
\tilde{Q}\,\pdv{f}{\tilde{Q}}-\frac{2n}{4-3n}\,\tilde{\tau}_0\,\pdv{f}{\tilde{\tau}_0} + \frac{2-n}{4-3n}\,\tilde{\varepsilon}\,\pdv{f}{\tilde{\varepsilon}} = 0,
\ee
which provides the curve along which $f$ is constant in the new parameter space.

\section{Specific cases of turbulent flow of Newtonian fluid, rough and smooth channel}
\label{app:specific_Newt_turb}

\subsection{Complete turbulence, rough channel}
\label{sec:vonKarman}
For sufficiently large Reynolds numbers, the friction factor corresponding to fully developed turbulent flow in a circular channel can quite accurately be described by Von Kármán's formula \citep{Haaland1983}, which holds for rough channels in the limit of high Reynolds numbers:

\be
f = \left\{-2.0\log_{10}\left(\frac{\delta}{3.7}\right)\right\}^{-2}.
\label{eq:VonKarman}
\ee
This equation is a limit case of the Colebrook-White equation, valid for the complete turbulent regime ($Re>3500/\delta$) \citep{Moody1944}. In terms of $\tilde{\varepsilon}$, $\tilde{Q}$ and $Re$, $f$ becomes

\be
\tilde{f} = \left\{-2.0\log_{10}\left(\frac{1}{3.7}\,\frac{\pi}{4}\frac{\tilde{\varepsilon}Re}{\tilde{Q}}\right)\right\}^{-2}.
\ee
Differentiation to $Re$ results in

\be
\pdv{\tilde{f}}{Re} = -\frac{\tilde{f}}{Re}\frac{2}{\ln\left(\frac{1}{3.7}\,\frac{\pi}{4}\frac{\tilde{\varepsilon}Re}{\tilde{Q}}\right) }
\ee
resulting in
\be
\tilde{B} = \tilde{f}\left(5- \frac{2}{\ln\left(\frac{1}{3.7}\,\frac{\pi}{4}\frac{\tilde{\varepsilon}Re}{\tilde{Q}}\right) } \right)
\ee

Contour plots of the optimal $Re$ as function $\tilde{Q}$ and $\tilde{\varepsilon}$ are presented in \fig[a]{fig:figure_3}, for the high-$\tilde{\varepsilon}$ limit.

\subsection{Turbulence, smooth channel}

\begin{figure}
\centering
\includegraphics[width=1\textwidth]{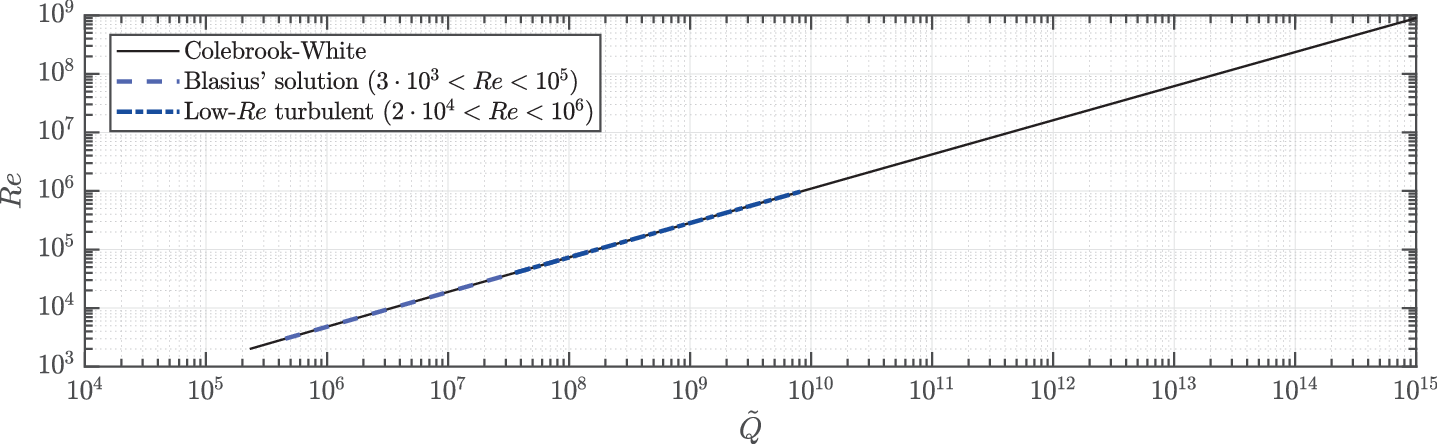} 
\caption{Optimal $Re$ as function of $\tilde{Q}$ for a Newtonian fluid in a smooth-wall channel according to the Colebrook-White equation, together with the Blasius' formula and a low-$Re$ turbulent approximation.}
\label{fig:optimization_specific_Newt}
\end{figure}

For a smooth-wall channel ($\varepsilon \rightarrow 0$), the Colebrook-White equation \citep{Colebrook1939} reduces to 

\be
\tilde{f} = \left\{-2.0\log_{10}\left(\frac{2.51}{Re\sqrt{\tilde{f}}}\right)\right\}^{-2}.
\label{eq:Colebrook_White_smooth}
\ee
Differentiation to $Re$ gives:

\be
\pdv{\tilde{f}}{Re} = \frac{\tilde{f}}{Re}\frac{2}{\ln\left(\frac{2.51}{Re\sqrt{\tilde{f}}}\right)-1}
\ee
resulting in

\be
\tilde{B} = \tilde{f}\left(5+\frac{2}{\ln\left(\frac{2.51}{Re\sqrt{\tilde{f}}}\right)-1}\right).
\ee
\Fig{fig:optimization_specific_Newt} shows the optimisation curve for $Re$ as function of $\tilde{Q}$. This result corresponds to the contour plot of the optimal $Re$ as function $\tilde{Q}$ and $\tilde{\varepsilon}$ as presented in \fig[a]{fig:figure_3} for $\tilde{\varepsilon}\rightarrow 0$.

\subsection{Low Reynolds number turbulence, smooth channel}
\label{app:Blasius}
Power-law approximations for the friction factor have been established for turbulent flows in smooth channels ($\varepsilon \rightarrow 0$), for specific and limited ranges of the Reynolds number, in the general form:

\be
\tilde{f} = \frac{c_1}{Re^{c_2}}
\label{eq:approximation_Newt_gen}
\ee
with $c_1$ and $c_2$ being two relation-dependent constants. This is an explicit expression for $\tilde{f}$ as function of $Re$, resulting in $\tilde{B}$ being:
\be
\tilde{B} = (5-c_2)\tilde{f} = \frac{c_1(5-c_2)}{Re^{c_2}}
\ee
For Newtonian fluids, the optimisation condition (\eq{eq:main_optimization_condition}) reduces to
\be
\tilde{B} \, Re^7=\frac{1024}{\pi^4}\, \tilde{Q}^4,
\ee
which can for this case be rewritten to
\be
Re =\left(\frac{1024}{c_1(5-c_2)}\right)^{\frac{1}{7-c_2}}\left(\frac{\tilde{Q}}{\pi}\right)^{\frac{4}{7-c_2}}.
\ee

Blasius' formula \citep{Blasius1912, Blasius1913}, which is valid for $3\times 10^3<Re<10^5$, is given by
\be
\tilde{f}=\frac{0.3164}{Re^{\frac{1}{4}}},
\label{eq: Blasius formula}
\ee
resulting in the optimisation condition
\be
Re = 2.629\left(\frac{\tilde{Q}}{\pi}\right)^{\frac{16}{27}}.
\ee

An empirical relation valid for turbulent flows at intermediate Reynolds numbers ($2\times 10^4 < Re <10^6$) \citep{Bejan2013_convectionBook} is given by
\be
\tilde{f}=\frac{0.184}{Re^{\frac{1}{5}}},
\label{eq:empirical_approx_Newt}
\ee
resulting in the optimisation condition
\be
Re = 2.822\left(\frac{\tilde{Q}}{\pi}\right)^{\frac{10}{17}}.
\ee

These explicit expressions for the optimal $Re$ as function of $\tilde{Q}$ are shown in \fig{fig:optimization_specific_Newt} for Blasius' formula (\eq{eq: Blasius formula}) and low $Re$-number turbulence (\eq{eq:empirical_approx_Newt}), revealing perfect coincidence with the more general Colebrook-White solution for smooth channels.

\section{Laminar and low Reynolds number turbulent flow of Bingham fluid, smooth channel}
\label{sec:Darby}

Considering a turbulent flow of a Bingham fluid in a circular pipe brings to empirical and numerical approximations for the friction factor $f$. Darby \textit{et al.} \citep{Darby1992} developed an empirical curve-fit equation, composed of the following structure:
\be
\begin{split}
&f = \left(f_L^{\chi_1}+f_T^{\chi_1}\right)^{1/{\chi_1}}\\
&f_T = 4\times 10^{\chi_2}Re^{-0.193}\\
&\chi_1 = 1.7+\frac{4\times 10^5}{Re}\\
&\chi_2 = -1.47\left(1+0.146\exp(-2.9\times 10^{-5}He)\right).
\label{eq:Darby}
\end{split}
\ee
where $f_L$ is the laminar friction factor as determined from the Buckingham-Reiner equation \citep{Reiner1926} (\eqsss{eq:main_phi_inReHe}{eq:f_laminar}{eq:Turb_psi_bingham} combined).

In the new set of dimensionless numbers, the friction factor $\tilde{f}$ becomes:
\be
\begin{split}
&\tilde{f} = \left(\tilde{f}_L^{\tilde{\chi}_1}+\tilde{f}_T^{\tilde{\chi}_1}\right)^{1/\tilde{\chi}_1}\\
&\tilde{f}_T = 4\times 10^{\tilde{\chi}_2}Re^{-0.193}\\
&\tilde{\chi}_1 = 1.7+\frac{4\times 10^5}{Re}\\
&\tilde{\chi}_2 = -1.47\left(1+0.146\exp(-2.9\times 10^{-5}\frac{16}{\pi^2}\frac{\tilde{Q}^2}{Re^2}\tilde{\tau}_0)\right).
\label{eq:Darby_newdim}
\end{split}
\ee

Implicit differentiation of $\tilde{f}$ to $Re$ results in the following expression

\begin{multline}
\frac{\partial \tilde{f}}{\partial Re} = \frac{\tilde{f}}{Re} \Bigg[ \frac{1.7-\tilde{\chi}_1}{\tilde{\chi}_1 }\left( \left(\frac{\tilde{f}_L}{\tilde{f}}\right)^{\tilde{\chi}_1} \ln \tilde{f}_L +\left(\frac{\tilde{f}_T}{\tilde{f}}\right)^{\tilde{\chi}_1} \ln \tilde{f}_T -\ln \tilde{f} \right) \\+ (J(\phi)-5)\left(\frac{\tilde{f}_L}{\tilde{f}}\right)^{\tilde{\chi}_1} +\left(\ln(10)Re\frac{\partial \tilde{\chi}_2}{\partial Re}-0.193\right)\left(\frac{\tilde{f}_T}{\tilde{f}}\right)^{\tilde{\chi}_1} \Bigg].
\end{multline}
where
\be
\frac{\partial \tilde{\chi}_2}{\partial Re} = -1.99\times 10^{-4}\times\frac{\tilde{Q}^2\tilde{\tau}_0}{\pi^2 Re^3}\exp\left(-4.64\times 10^{-4}\frac{\tilde{Q}^2\tilde{\tau}_0}{\pi^2 Re^2}\right)
\ee
resulting in
\be
\tilde{B} = \tilde{f}\left(5+\frac{Re}{\tilde{f}}\pdv{\tilde{f}}{Re}\right).
\ee

\begin{figure}[h]
\begin{center}
   \centering
   \includegraphics[width=1\textwidth]{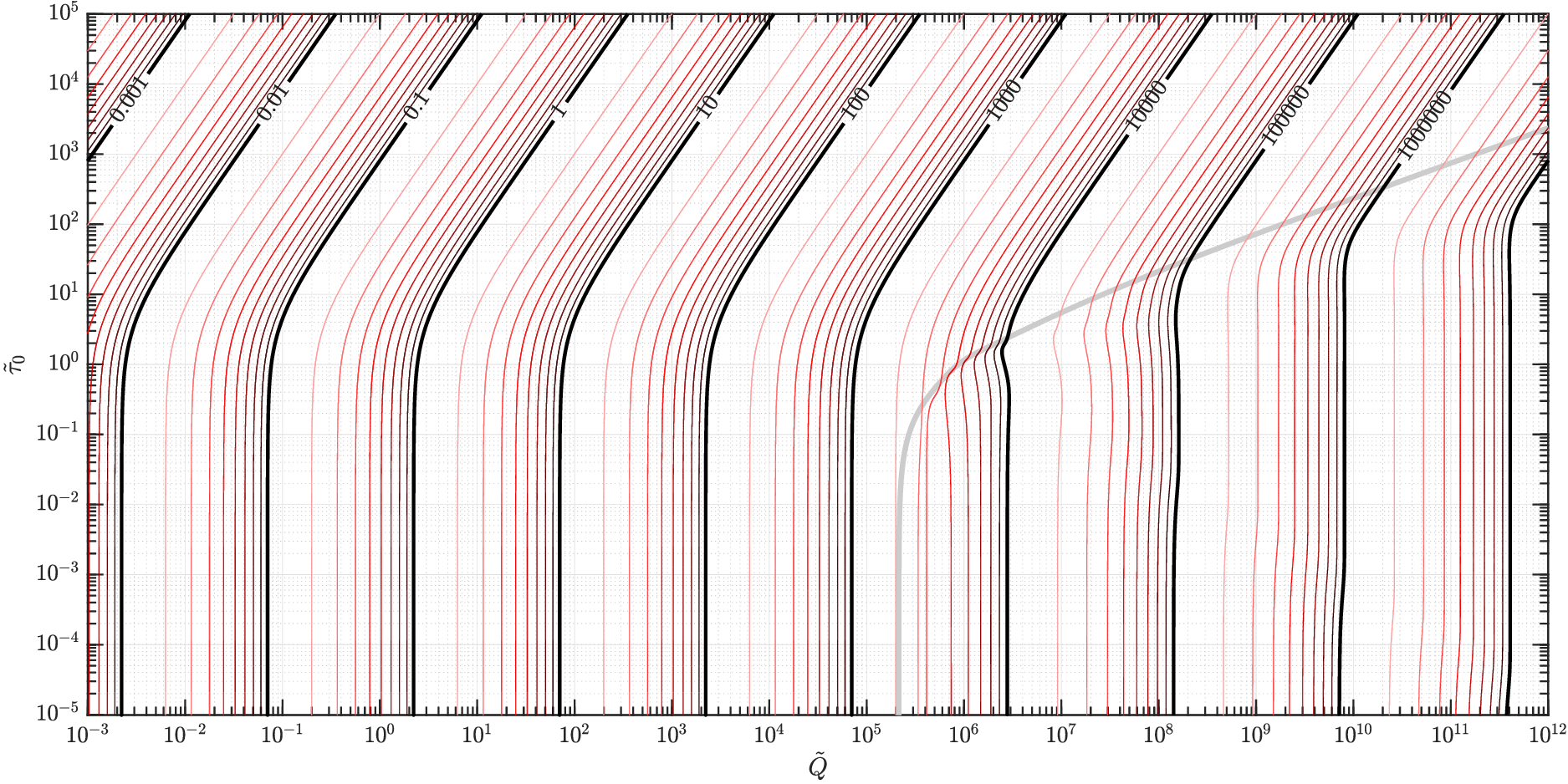} 
\caption{Contour plot of the optimal Reynolds number as a function of $\tilde{Q}$ and $\tilde{\tau}_0$, for both laminar and low Reynolds number turbulent flow of a Bingham fluid (\eq{eq:Darby}). The red-scale contour lines represent the Reynolds numbers $2\times 10^x$, $3\times 10^x$, ... $9\times 10^x$ with decreasing brightness. The thick grey line represents the critical Reynolds number.}
\label{fig:Darby}
\end{center}
\end{figure}

The set of equations for this fluid model is relatively complex, but has as benefit that it covers both the laminar and turbulent regimes. The contour plot in \fig{fig:Darby} gives the optimal Reynolds number in a channel for given $\tilde{\tau}_0$ and $\tilde{Q}$.

\section{Approximation of $x$ for different fluid models}
\label{app:deriv_x}
In network optimisation, many attempts have been made to find the proportionality between the flow rate $Q$ and the channel radius $R$ for optimised networks, in the following form:
\be
Q\propto R^x
\label{eq:propQ_R_app}
\ee
In the optimisation method of the present study, only $Re$ and $\tilde{Q}$ contain parameters involving $R$ and $Q$, whereas $\tilde{\tau}_0$, $\tilde{\varepsilon}$ and $n$ are independent of $R$ and $Q$. Therefore, when knowing the proportionality between $\tilde{Q}$ and $Re$, one can obtain $x$ using \eq{eq:propQ_R_app}. However, this is only posible if $\tilde{B}\propto Re^{c_1}\tilde{Q}^{c_2}$. However, in most of the cases (except for e.g. Blasius' formula), this is not true. Therefore, $x$ is calculated by locally approximating $\tilde{B}$ by a power function using

\be
x = \frac{R}{Q}\frac{\partial Q}{\partial R}= \frac{(4-3n)\frac{Re}{\tilde{Q}}\pdv{\tilde{Q}}{Re}}{(2-n)\frac{Re}{\tilde{Q}}\pdv{\tilde{Q}}{Re}-1} = 3-\frac{2\frac{Re}{\tilde{Q}}\pdv{\tilde{Q}}{Re}-3}{(2-n)\frac{Re}{\tilde{Q}}\pdv{\tilde{Q}}{Re}-1},
\label{eq:x_def}
\ee
where $\frac{Re}{\tilde{Q}}\pdv{\tilde{Q}}{Re}$ is derived to be:

\noindent\scalebox{1}{
\begin{minipage}{\linewidth} 
\begin{multline}
\frac{Re}{\tilde{Q}}\pdv{\tilde{Q}}{Re} = - \Bigg[\left((4-3n)\frac{Re}{\tilde{B}}\left(\pdv{\tilde{B}}{Re}\right)_{\tilde{f},\tilde{Q},\phi} -4\frac{\phi}{\tilde{B}}\left(\pdv{\tilde{B}}{\phi}\right)_{\tilde{f},\tilde{Q},Re} +7\right)\left(1+\frac{\phi}{\tilde{f}}\left(\pdv{\tilde{f}}{\phi}\right)_{\tilde{Q},Re} \right)+\\+\left(\frac{\tilde{f}}{\tilde{B}}\left(\pdv{\tilde{B}}{\tilde{f}}\right)_{\tilde{Q},\phi,Re}-\frac{\phi}{\tilde{B}}\left(\pdv{\tilde{B}}{\phi}\right)_{\tilde{f},\tilde{Q},Re}\right)\left( (4-3n)\frac{Re}{\tilde{f}}\left(\pdv{\tilde{f}}{Re}\right)_{\tilde{Q},\phi}-4\frac{\phi}{\tilde{f}}\left(\pdv{\tilde{f}}{\phi}\right)_{\tilde{Q},Re} \right)\Bigg]\times\\ \Bigg[ \left(2n\frac{\phi}{\tilde{B}}\left(\pdv{\tilde{B}}{\phi}\right)_{\tilde{f},\tilde{Q},Re}+(4-3n)\frac{\tilde{Q}}{\tilde{B}}\left(\pdv{\tilde{B}}{\tilde{Q}}\right)_{\tilde{f},\phi,Re}-2(n+1)\right)\left(1+\frac{\phi}{\tilde{f}}\left(\pdv{\tilde{f}}{\phi}\right)_{\tilde{Q},Re} \right)+\\+\left(\frac{\tilde{f}}{\tilde{B}}\left(\pdv{\tilde{B}}{\tilde{f}}\right)_{\tilde{Q},\phi,Re}-\frac{\phi}{\tilde{B}}\left(\pdv{\tilde{B}}{\phi}\right)_{\tilde{f},\tilde{Q},Re}\right)\left( 2n\frac{\phi}{\tilde{f}}\left(\pdv{\tilde{f}}{\phi}\right)_{\tilde{Q},Re}+(4-3n)\frac{\tilde{Q}}{\tilde{f}}\left(\pdv{\tilde{f}}{\tilde{Q}}\right)_{\phi,Re} \right)     \Bigg]^{-1}.
\label{eq:domega_dRe}
\end{multline}
\end{minipage}
}

\Eq{eq:domega_dRe} is an expression containing 7 differentials  of $\tilde{B}$ or $\tilde{f}$ to $Re$, $\phi$, $\tilde{f}$ or $\tilde{Q}$. All differentials can be calculated with keeping the parameters in subscript fixed. The differentials have to be calculated for every flow model separately. The calculated value for $x$ is only valid around its corresponding point in the ($Re,\tilde{\tau}_0,\tilde{\varepsilon},n$)-space with a limited range depending on the change in $x$. Expressions for the differentials for the fluid models presented in the following sections. 

\subsection{Laminar flow of a Newtonian, power-law, Bingham and Herschel-Bulkley fluid}

For a laminar flow, the friction factor was found to be
\be
\tilde{f} =\frac{64}{Re \,\psi(\phi,n)^n},
\ee
resulting in an expression for $\tilde{B}$ being
\be
\tilde{B} = \tilde{f}J(\phi,n),
\ee
where $J(\phi,n)$ is defined as
\be
J\equiv 1+\frac{3n}{1-\frac{n\phi}{\psi}\pdv{\psi}{\phi}}.
\ee

Differentiation provides the needed differentials in \eq{eq:domega_dRe}, resulting in the expression for $x$ (\eq{eq:x_def}):

\be
\begin{aligned}
\frac{\tilde{f}}{\tilde{B}}\left(\pdv{\tilde{B}}{\tilde{f}}\right)_{\tilde{Q},\phi,Re}&=1,\quad \quad
\frac{Re}{\tilde{B}}\left(\pdv{\tilde{B}}{Re}\right)_{\tilde{f},\tilde{Q},\phi}&=0, \quad\quad
\frac{\phi}{\tilde{B}}\left(\pdv{\tilde{B}}{\phi}\right)_{\tilde{f},\tilde{Q},Re}&=\frac{\phi}{J}\pdv{J}{\phi},\\
\frac{\tilde{Q}}{\tilde{B}}\left(\pdv{\tilde{B}}{\tilde{Q}}\right)_{\tilde{f},\phi,Re}&=0,\quad\quad
\frac{Re}{\tilde{f}}\left(\pdv{\tilde{f}}{Re}\right)_{\tilde{Q},\phi}&=-1,\quad\quad
\frac{\phi}{\tilde{f}}\left(\pdv{\tilde{f}}{\phi}\right)_{\tilde{Q},Re}&=-\frac{n\phi}{\psi}\pdv{\psi}{\phi},\\
\frac{\tilde{Q}}{\tilde{f}}\left(\pdv{\tilde{f}}{\tilde{Q}}\right)_{\phi,Re}&=0.
\end{aligned}
\ee

Substituting everything into \eq{eq:domega_dRe} results in $\frac{Re}{\tilde{Q}}\pdv{\tilde{Q}}{Re} = \frac{3}{2}$, resulting in $x=3$. This holds for laminar flow of all treated fluid models.

\subsection{Turbulent flow of Newtonian fluid, rough and smooth channel}

A turbulent flow of a Newtonian fluid in a rough or smooth channel can be described by the Colebrook-White equation \citep{Colebrook1939}, which is in the applied nondimensionalisation:

\be
\tilde{f} = \left\{-2.0\log_{10}\left(k_1+k_2\right)\right\}^{-2}.
\label{eq:Colebrook_White_SI}
\ee
with
\begin{equation}
k_1 \equiv \frac{1}{3.7}\,\frac{\pi}{4}\frac{\tilde{\varepsilon}Re}{\tilde{Q}}
\end{equation}
\begin{equation}
k_2 \equiv \frac{2.51}{\sqrt{\tilde{f}}Re}
\end{equation}
Differentiation provides the needed differentials in \eq{eq:domega_dRe}:

\be
\begin{aligned}
\frac{\tilde{f}}{\tilde{B}}\left(\pdv{\tilde{B}}{\tilde{f}}\right)_{\tilde{Q},\phi,Re}&=1-\frac{\tilde{f}}{\tilde{B}}\left[  \frac{2k_1 k_2\ln\left(k_1+k_2 \right)-k_2^2}{\left[\left(k_1+k_2\right)\ln\left(k_1+k_2\right)-k_2\right]^{2}}\right],\\
\frac{Re}{\tilde{B}}\left(\pdv{\tilde{B}}{Re}\right)_{\tilde{f},\tilde{Q},\phi}&=\frac{\tilde{f}}{\tilde{B}}\left[ \frac{-8k_1k_2\ln(k_1+k_2)+2k_1^2+2k_2^2}{[(k_1+k_2)\ln(k_1+k_2)-k_2]^2} \right],\quad
&\frac{\phi}{\tilde{B}}\left(\pdv{\tilde{B}}{\phi}\right)_{\tilde{f},\tilde{Q},Re}=0,\\
\frac{\tilde{Q}}{\tilde{B}}\left(\pdv{\tilde{B}}{\tilde{Q}}\right)_{\tilde{f},\phi,Re}&=\frac{ \tilde{f}}{\tilde{B}}\left[  \frac{4k_1k_2\ln(k_1+k_2)-2k_1^2}{[(k_1+k_2)\ln(k_1+k_2)-k_2]^2}  \right],\\
\frac{Re}{\tilde{f}}\left(\pdv{\tilde{f}}{Re}\right)_{\tilde{Q},\phi}&=\frac{2(k_2-k_1)}{(k_1+k_2)\ln(k_1+k_2)-k_2},\quad\quad
&\frac{\phi}{\tilde{f}}\left(\pdv{\tilde{f}}{\phi}\right)_{\tilde{Q},Re}=0,\\
\frac{\tilde{Q}}{\tilde{f}}\left(\pdv{\tilde{f}}{\tilde{Q}}\right)_{\phi,Re}&=\frac{2k_1}{(k_1+k_2)\ln(k_1+k_2)-k_2}. 
\end{aligned}
\ee

\subsubsection{Complete turbulent flow of Newtonian fluid, rough channel}

For sufficiently large Reynolds numbers, the friction factor corresponding to complete turbulence in a circular channel can accurately be described by Von Kármán's formula \citep{Haaland1983}:
\be
f = \left\{-2.0\log_{10}\left(\frac{1}{3.7}\,\frac{\pi}{4}\frac{\tilde{\varepsilon}Re}{\tilde{Q}}\right)\right\}^{-2}.
\ee
Differentiation provides the needed differentials in \eq{eq:domega_dRe}:

\be
\begin{aligned}
\frac{\tilde{f}}{\tilde{B}}\left(\pdv{\tilde{B}}{\tilde{f}}\right)_{\tilde{Q},\phi,Re}&=1,\quad\quad
\frac{Re}{\tilde{B}}\left(\pdv{\tilde{B}}{Re}\right)_{\tilde{f},\tilde{Q},\phi}=\frac{\tilde{f}}{\tilde{B}}\frac{2}{\left(\ln\left(\frac{1}{3.7}\,\frac{\pi}{4}\frac{\tilde{\varepsilon}Re}{\tilde{Q}}\right)\right)^2},\\
\frac{\phi}{\tilde{B}}\left(\pdv{\tilde{B}}{\phi}\right)_{\tilde{f},\tilde{Q},Re}&=0,\quad\quad
\frac{\tilde{Q}}{\tilde{B}}\left(\pdv{\tilde{B}}{\tilde{Q}}\right)_{\tilde{f},\phi,Re}=\frac{\tilde{f}}{\tilde{B}}\frac{-2}{\left(\ln\left(\frac{1}{3.7}\,\frac{\pi}{4}\frac{\tilde{\varepsilon}Re}{\tilde{Q}}\right)\right)^2},\\
\frac{Re}{\tilde{f}}\left(\pdv{\tilde{f}}{Re}\right)_{\tilde{Q},\phi}&=\frac{-1}{\ln\left(\frac{1}{3.7}\,\frac{\pi}{4}\frac{\tilde{\varepsilon}Re}{\tilde{Q}}\right)},\quad\quad
\frac{\phi}{\tilde{f}}\left(\pdv{\tilde{f}}{\phi}\right)_{\tilde{Q},Re}=0,\\
\frac{\tilde{Q}}{\tilde{f}}\left(\pdv{\tilde{f}}{\tilde{Q}}\right)_{\phi,Re}&=\frac{1}{\ln\left(\frac{1}{3.7}\,\frac{\pi}{4}\frac{\tilde{\varepsilon}Re}{\tilde{Q}}\right)}.
\end{aligned}
\ee

\subsubsection{Turbulence, smooth channel}

For a smooth-wall channel ($\varepsilon \rightarrow 0$), the Colebrook-White equation \citep{Colebrook1939} reduces to 

\be
\tilde{f} = \left\{-2.0\log_{10}\left(\frac{2.51}{Re\sqrt{\tilde{f}}}\right)\right\}^{-2}.
\ee
Differentiation provides the needed differentials in \eq{eq:domega_dRe}:
\be
\begin{aligned}
\frac{\tilde{f}}{\tilde{B}}\left(\pdv{\tilde{B}}{\tilde{f}}\right)_{\tilde{Q},\phi,Re}&=1+\frac{\tilde{f}}{\tilde{B}}\left[  \frac{1}{\left[\ln\left(\frac{2.51}{\sqrt{\tilde{f}}Re}\right)-1\right]^{2}}\right],\\
\frac{Re}{\tilde{B}}\left(\pdv{\tilde{B}}{Re}\right)_{\tilde{f},\tilde{Q},\phi}&=\frac{\tilde{f}}{\tilde{B}}\left[ \frac{2}{[\ln(\frac{2.51}{\sqrt{\tilde{f}}Re})-1]^2} \right],\quad\quad
\frac{\phi}{\tilde{B}}\left(\pdv{\tilde{B}}{\phi}\right)_{\tilde{f},\tilde{Q},Re}=0,\\
\frac{\tilde{Q}}{\tilde{B}}\left(\pdv{\tilde{B}}{\tilde{Q}}\right)_{\tilde{f},\phi,Re}&=0,\quad\quad
\frac{Re}{\tilde{f}}\left(\pdv{\tilde{f}}{Re}\right)_{\tilde{Q},\phi}=\frac{2}{\ln(\frac{2.51}{\sqrt{\tilde{f}}Re})-1},\\
\frac{\phi}{\tilde{f}}\left(\pdv{\tilde{f}}{\phi}\right)_{\tilde{Q},Re}&=0,\quad\quad
\frac{\tilde{Q}}{\tilde{f}}\left(\pdv{\tilde{f}}{\tilde{Q}}\right)_{\phi,Re}=0. 
\end{aligned}
\ee

\subsubsection{Low Reynolds number turbulence, smooth channel}

For certain Reynolds number regimes of flow in a smooth channel ($\varepsilon\rightarrow 0$), approximations for the friction factor have been formulated, such as Blasius' formula and empirical relations. These are often in the following form \citep{Uylings1977}:

\be
\tilde{f} = \frac{c_1}{Re^{c_2}}
\ee
Differentiation provides the needed differentials in \eq{eq:domega_dRe}:

\be
\begin{aligned}
\frac{\tilde{f}}{\tilde{B}}\left(\pdv{\tilde{B}}{\tilde{f}}\right)_{\tilde{Q},\phi,Re}&=1,\quad\quad
\frac{Re}{\tilde{B}}\left(\pdv{\tilde{B}}{Re}\right)_{\tilde{f},\tilde{Q},\phi}=0,\quad\quad
\frac{\phi}{\tilde{B}}\left(\pdv{\tilde{B}}{\phi}\right)_{\tilde{f},\tilde{Q},Re}=0,\\
\frac{\tilde{Q}}{\tilde{B}}\left(\pdv{\tilde{B}}{\tilde{Q}}\right)_{\tilde{f},\phi,Re}&=0,\quad\quad
\frac{Re}{\tilde{f}}\left(\pdv{\tilde{f}}{Re}\right)_{\tilde{Q},\phi}=-c_2,\quad\quad
\frac{\phi}{\tilde{f}}\left(\pdv{\tilde{f}}{\phi}\right)_{\tilde{Q},Re}=0,\\
\frac{\tilde{Q}}{\tilde{f}}\left(\pdv{\tilde{f}}{\tilde{Q}}\right)_{\phi,Re}&=0.
\end{aligned}
\ee

Substituting everything into \eq{eq:domega_dRe} results in $\frac{Re}{\tilde{Q}}\pdv{\tilde{Q}}{Re} = \frac{7-c_2}{4}$, resulting in $x=\frac{7-c_2}{3-c_2}$. For Blasius' equation (\eq{eq: Blasius formula}), $x$ reduces to $x = \frac{27}{11}$ and for the empirical relation (\eq{eq:empirical_approx_Newt}) $x$ reduces to $x=\frac{17}{7}$.

\subsection{Turbulent flow of non-Newtonian fluids, smooth channel}
\subsubsection{Power-law fluid}
Turbulent pipe flow of non-Newtonian power-law fluids at low Reynolds numbers in a smooth circular channel ($\varepsilon\rightarrow 0$) described by Dodge and Metzner \citep{Dodge1959} is given in the following implicit relation for the Darcy friction factor $f$:
\be
\frac{2}{\sqrt{f}}=\frac{4}{n^{0.75}}\log_{10}\left(Re\left(\frac{f}{4}\right)^{1-n/2}\right)-\frac{0.4}{n^{1.2}}
\ee
Differentiation provides the needed differentials in \eq{eq:domega_dRe}:
\be
\begin{aligned}
\frac{\tilde{f}}{\tilde{B}}\left(\pdv{\tilde{B}}{\tilde{f}}\right)_{\tilde{Q},\phi,Re}&=1-\frac{\tilde{f}}{\tilde{B}}(4-3n)\left(\frac{n^{0.75}\ln(10)}{4\sqrt{\tilde{f}}}+1-\frac{n}{2}\right)^{-2} \frac{n^{0.75}\ln(10)}{8\sqrt{\tilde{f}}},\\
\frac{Re}{\tilde{B}}\left(\pdv{\tilde{B}}{Re}\right)_{\tilde{f},\tilde{Q},\phi}&=0,\quad\quad
\frac{\phi}{\tilde{B}}\left(\pdv{\tilde{B}}{\phi}\right)_{\tilde{f},\tilde{Q},Re}=0,\quad\quad
\frac{\tilde{Q}}{\tilde{B}}\left(\pdv{\tilde{B}}{\tilde{Q}}\right)_{\tilde{f},\phi,Re}=0,\\
\frac{Re}{\tilde{f}}\left(\pdv{\tilde{f}}{Re}\right)_{\tilde{Q},\phi}&=-\left(\frac{n^{0.75}\ln(10)}{4\sqrt{\tilde{f}}}+1-\frac{n}{2}\right)^{-1},\\
\frac{\phi}{\tilde{f}}\left(\pdv{\tilde{f}}{\phi}\right)_{\tilde{Q},Re}&=0,\quad\quad
\frac{\tilde{Q}}{\tilde{f}}\left(\pdv{\tilde{f}}{\tilde{Q}}\right)_{\phi,Re}=0.
\end{aligned}
\ee

\subsubsection{Herschel-Bulkley fluid}
For a turbulent flow of a Herschel-Bulkley fluid in a smooth circular channel ($\varepsilon \rightarrow 0$), Torrance \citep{Garcia1986} developed a relationship for the Darcy friction factor:

\be
\frac{2}{\sqrt{\tilde{f}}}=0.45-\frac{2.75}{n}+\frac{1.97}{n}\ln(1-\phi) +\frac{1.97}{n}\ln\left(Re\left(\frac{3n+1}{4n}\right)^n \left(\frac{\tilde{f}}{4}\right)^{1-\frac{n}{2}}\right).
\ee

Differentiation provides the needed differentials in \eq{eq:domega_dRe}:
\be
\begin{aligned}
\frac{\tilde{f}}{\tilde{B}}\left(\pdv{\tilde{B}}{\tilde{f}}\right)_{\tilde{Q},\phi,Re}&=1-\frac{\tilde{f}}{\tilde{B}}\frac{4-3n+\frac{4\phi}{1-\phi}}{\left( \frac{n}{1.97\sqrt{\tilde{f}}}+\frac{\phi}{1-\phi}+1-\frac{n}{2} \right)}\frac{\frac{1}{2}n}{1.97\sqrt{\tilde{f}}},\\
\frac{Re}{\tilde{B}}\left(\pdv{\tilde{B}}{Re}\right)_{\tilde{f},\tilde{Q},\phi}&=0,\quad\quad
\frac{\phi}{\tilde{B}}\left(\pdv{\tilde{B}}{\phi}\right)_{\tilde{f},\tilde{Q},Re}=-\frac{ \tilde{f}}{\tilde{B}} \frac{\left(\frac{4n}{1.97\sqrt{\tilde{f}}}+n\right)\frac{\phi}{(1-\phi)^2} }{\left(\frac{n}{1.97\sqrt{\tilde{f}}}+\frac{\phi}{1-\phi}+1-\frac{n}{2}\right)^2},\\
\frac{\tilde{Q}}{\tilde{B}}\left(\pdv{\tilde{B}}{\tilde{Q}}\right)_{\tilde{f},\phi,Re}&=0,\quad\quad
\frac{Re}{\tilde{f}}\left(\pdv{\tilde{f}}{Re}\right)_{\tilde{Q},\phi}=\frac{-1}{\frac{n}{1.97\sqrt{\tilde{f}}}+1-\frac{n}{2}},\\
\frac{\phi}{\tilde{f}}\left(\pdv{\tilde{f}}{\phi}\right)_{\tilde{Q},Re}&=\frac{\frac{\phi}{1-\phi}}{\frac{n}{1.97\sqrt{\tilde{f}}}+1-\frac{n}{2}},\quad\quad
\frac{\tilde{Q}}{\tilde{f}}\left(\pdv{\tilde{f}}{\tilde{Q}}\right)_{\phi,Re}=0.
\end{aligned}
\ee

\section{Extra figures for Herschel-Bulkley fluids ($n=0.5$ and $n=1.5$)}
\label{app:extra_figures_HB}

\begin{figure}[h!]
\centering
\includegraphics[width=1\textwidth]{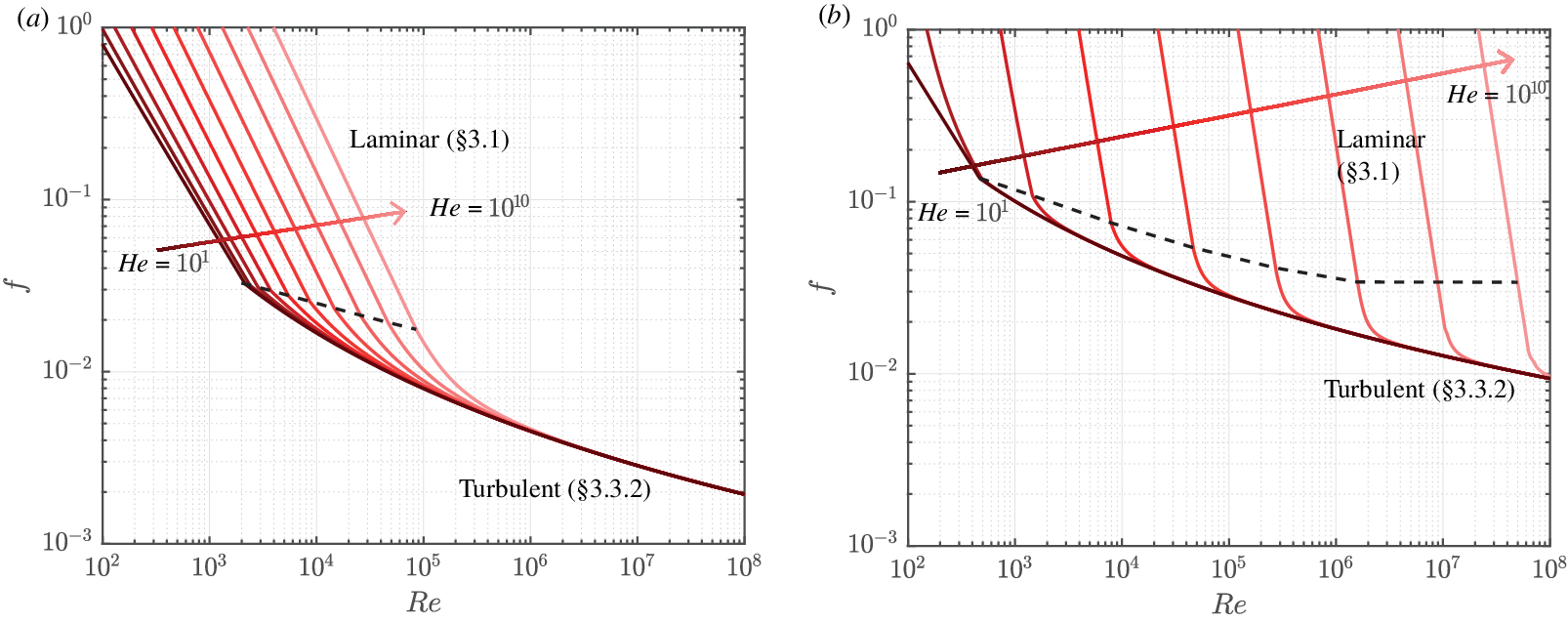} 
\caption{Friction factor as function of the Reynolds number for Herschel-Bulkley fluids. The dashed black line indicates the transition from laminar to turbulent flow. The red-scale lines represent constant values of the Hedstr\"om number from $10^{1}, 10^2, ..., 10^{10}$ with decreasing brightness. \scnr{a} $n=0.5$. \scnr{b} $n=1.5$.}
\label{fig:f_Re_plots_HB}
\end{figure}

\begin{figure}[p]
\begin{center}
   \centering
   \includegraphics[width=1\textwidth]{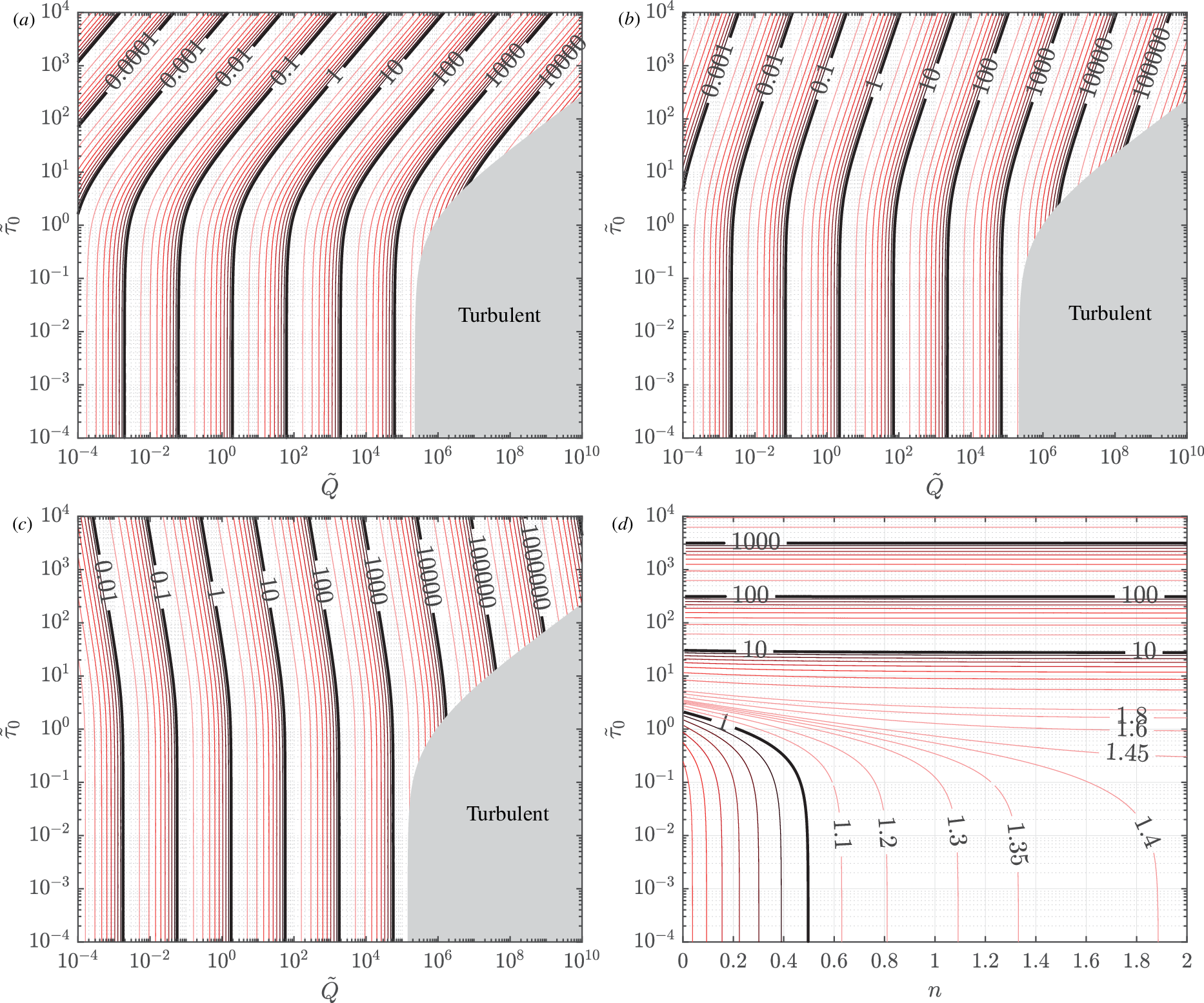} 
\caption{Contour plots of the optimisation condition for a laminar flow of a Herschel-Bulkley fluid (\eq{eq:laminar_opt_Re}). The red-scale contour lines represent values $2\times 10^x$, $3\times 10^x$, ... $9\times 10^x$ with decreasing brightness. \scnr{a} Optimal Reynolds number as a function of $\tilde{\tau}_0$ and $\tilde{Q}$ for a Herschel-Bulkley fluid with $n=0.5$. \scnr{b} Optimal Reynolds number as a function of $\tilde{\tau}_0$ and $\tilde{Q}$ for a Herschel-Bulkley fluid with $n=1.0$. \scnr{c} Optimal Reynolds number as a function of $\tilde{\tau}_0$ and $\tilde{Q}$ for a Herschel-Bulkley fluid with $n=1.5$. \scnr{d} Optimal $\tilde{R}^3/\tilde{Q}$ as function of $n$ and $\tilde{\tau}_0$. This optimisation plot covers all laminar flows of Herschel-Bulkley fluids.}
\label{fig:laminar_all}
\end{center}
\end{figure}

\begin{figure}[p]
\centering
   \includegraphics[width=\textwidth]{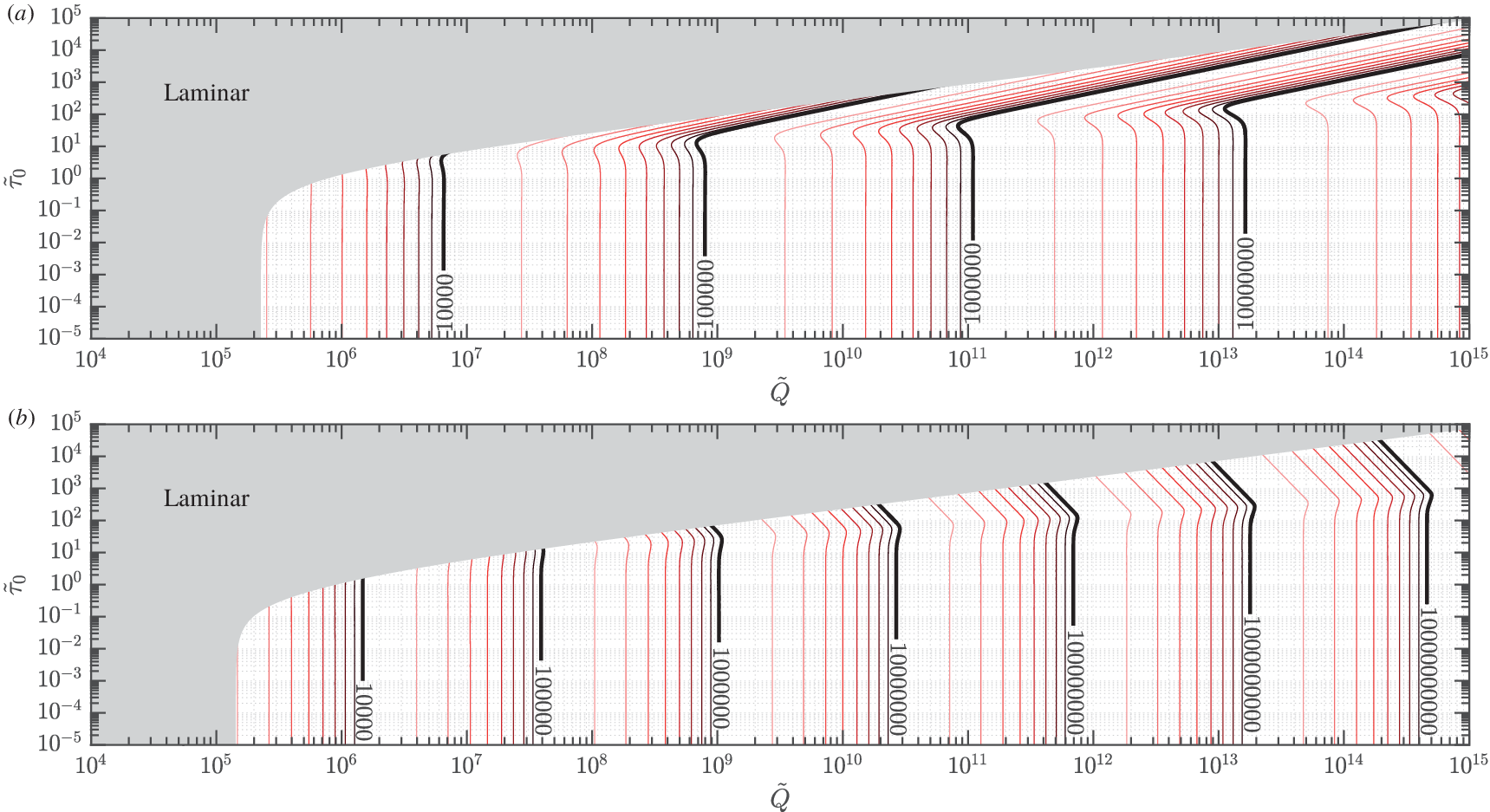} 
\caption{Contour plots of the optimal Reynolds number as a function of $\tilde{Q}$ and $\tilde{\tau}_0$ for a turbulent flow of a Herschel-Bulkley fluid (\eq{eq:Torrance}) at constant $n$. The red-scale contour lines represent the Reynolds numbers $2\times 10^x$, $3\times 10^x$, ... $9\times 10^x$ with decreasing brightness. \scnr{a} $n=0.5$. \scnr{b} $n=1.5$.}
\label{fig:contour_plots_HB}
\end{figure}

\begin{figure}[p]
\centering
   \includegraphics[width=\textwidth]{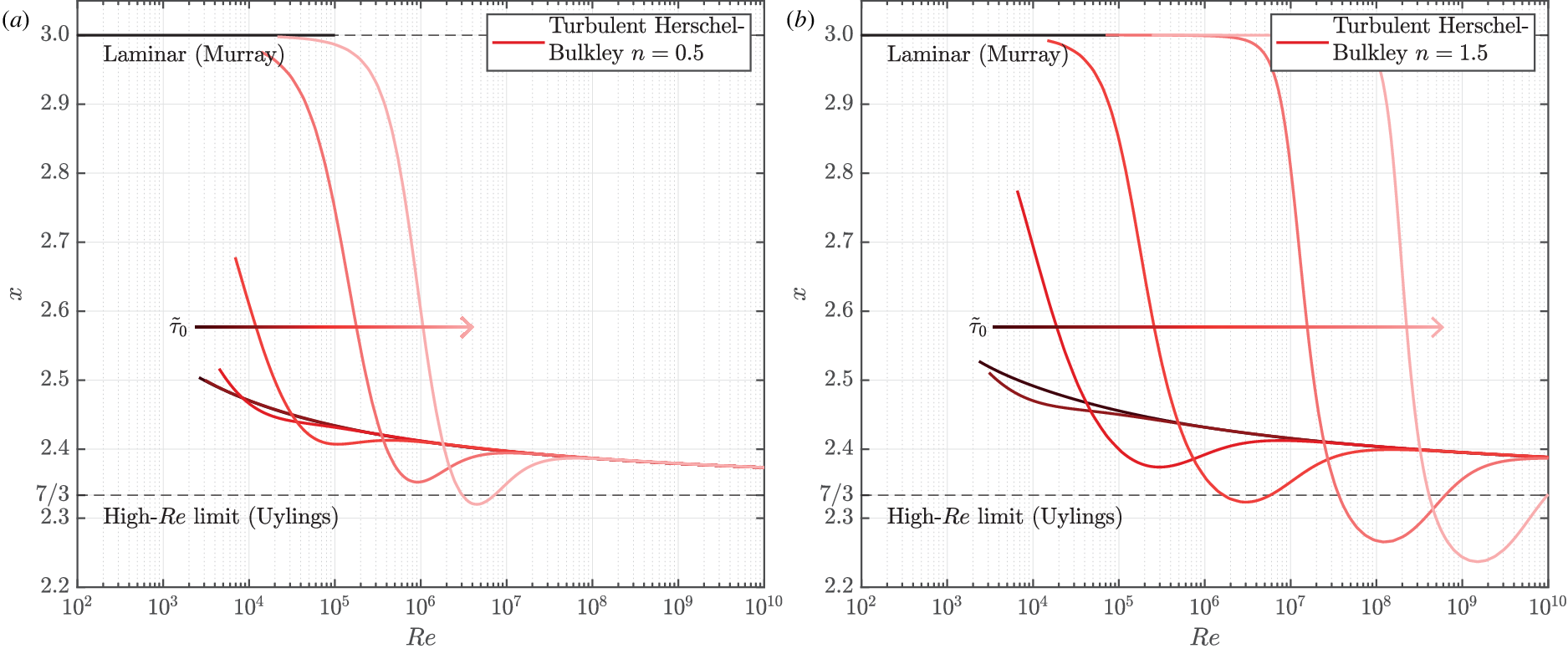} 
\caption{Plot of $x$ (\eq{eq:main_x_def}) as function of $Re$ for Turbulent flow of a Herschel-Bulkley fluid (Torrance) as discussed in \sec{sec:turb_HB}. The dashed lines for $x=3$ and $x=7/3$ show the expected limit cases for laminar flow and high-turbulent flow, respectively \citep{Uylings1977}. The different lines are for $\tilde{\tau}_0 \in [0.1,0.25,1,2.5,10,25]$. \scnr{a} $n= 0.5$. \scnr{b} $n= 1.5$.}
\label{fig:X_HB}
\end{figure}

\end{document}